\documentclass[a4paper,11pt]{article}
\usepackage{jcappub}

\usepackage{amsfonts,amssymb,amsmath}
\usepackage[dvipsnames, table]{xcolor}
\usepackage{epsfig}
\usepackage{bm}
\usepackage{natbib}
\usepackage{url}
\usepackage{dcolumn}
\usepackage{color}
\usepackage{graphicx}
\usepackage{comment}
\usepackage{hyperref}
\usepackage[normalem]{ulem}
\hypersetup{colorlinks   = true,
            urlcolor     = blue,
            citecolor    = blue,
            linkcolor    = blue,
            menucolor    = blue,
            anchorcolor  = blue,
            filecolor    = blue
}

\title{Dark photon dark matter from flattened axion potentials}

\author[a,1]{Hong-Yi Zhang,\note{Corresponding author.}}
\author[b]{ Paola Arias,}
\author[a]{ Andrew Cheek,}
\author[b]{ Enrico D.\ Schiappacasse,}
\author[c,d]{ Luca Visinelli,}
\author[e,f]{ and Leszek Roszkowski}
\affiliation[a]{Tsung-Dao Lee Institute \& School of Physics and Astronomy, Shanghai Jiao Tong University, Shanghai 201210, China}
\affiliation[b]{Facultad de Ingenier\'ia,  Universidad San Sebasti\'an, Bellavista 7, Santiago 8420524, Chile}
\affiliation[c]{Dipartimento di Fisica ``E.R.\ Caianiello'', Universit\`a degli Studi di Salerno, Via Giovanni Paolo II, 132 - 84084 Fisciano (SA), Italy}
\affiliation[d]{Istituto Nazionale di Fisica Nucleare - Gruppo Collegato di Salerno - Sezione di Napoli, Via Giovanni Paolo II, 132 - 84084 Fisciano (SA), Italy}
\affiliation[e]{Astrocent, Nicolaus Copernicus Astronomical Center Polish Academy of Sciences, ul.\ Rektorska 4, 00-614, Warsaw, Poland}
\affiliation[f]{National Centre for Nuclear Research, ul. Pasteura 7, 02-093 Warsaw, Poland}
\emailAdd{hongyi18@sjtu.edu.cn}
\emailAdd{paola.arias@uss.cl}
\emailAdd{acheek@sjtu.edu.cn}
\emailAdd{enrico.schiappacasse@uss.cl}
\emailAdd{lvisinelli@unisa.it}
\emailAdd{leszek.roszkowski@ncbj.gov.pl}

\abstract{Dark photons can be resonantly produced in the early universe via their coupling to an oscillating axion field. However, this mechanism typically requires large axion--dark photon couplings or some degree of fine-tuning. In this work, we present a new scenario in which efficient dark photon production arises from axion potentials that are shallower than quadratic at large field values. For moderately large initial misalignment angles, the oscillation of the axion field can trigger either efficient dark photon production or strong axion self-resonance via parametric resonance. When self-resonance dominates and disrupts the field's homogeneity, we show that oscillons---localized, oscillating axion field configurations---naturally form and can sustain continued dark photon production, provided the coupling is $\gtrsim \mathcal O(1)$. For dark photon mass up to three orders of magnitude below the axion mass, the produced dark photons can account for a significant fraction of the present-day dark matter. We support this scenario with numerical lattice simulations of a benchmark model. Our results further motivate experimental searches for ultralight dark photon dark matter. The simulation code is publicly available at \url{https://github.com/hongyi18/AxionDarkPhotonSimulator}.}

\begin{document}
\maketitle
\flushbottom

\section{Introduction}
The nature of dark matter (DM) remains one of the most pressing open questions in fundamental physics. Increasingly precise cosmological and astrophysical data, combined with null results from direct detection experiments, have motivated the exploration of alternative DM candidates beyond weakly interacting massive particles (WIMPs)~\cite{Bertone:2004pz, Feng:2010gw, Baer:2014eja, Cirelli:2024ssz} and axions---well-motivated pseudoscalar particles in particle physics extensions and string theory~\cite{Marsh:2015xka, DiLuzio:2020wdo, Chadha-Day:2021szb}. Among these, dark photons, hypothetical vector bosons that may or may not arise from an additional $U(1)$ gauge symmetry~\cite{Redondo:2008ec, Jaeckel:2010ni}, have emerged as compelling candidates, particularly in the ultralight mass regime $m_X \lesssim 10\,\mathrm{eV}$. In this regime, the wave nature of DM manifests on galactic and even cosmological scales, opening up new avenues to explain anomalies in structure formation and astrophysical dynamics. Ultralight DM has been proposed as a possible explanation for the diversity of galactic rotation curves~\cite{Oman:2015xda, Luu:2024lfq}, strong gravitational lensing anomalies~\cite{Amruth:2023xqj},\footnote{The validity of the strong lensing anomaly is currently under debate~\cite{2025arXiv250603132M}.} the final parsec problem in the dynamics of supermassive black hole mergers~\cite{Bromley:2023yfi, Koo:2023gfm}, and anomalous features in DM halo collisions~\cite{Koo:2025qac}.

Unlike scalar fields, vector DM possesses a nontrivial polarization structure, leading to qualitatively different behavior. One particularly interesting context in which the vector nature becomes crucial is the formation of solitons---coherent, localized DM configurations that may form in DM halos~\cite{Zhang:2021xxa, Jain:2021pnk, Zhang:2023ktk, Zhang:2024bjo, Amin:2022pzv, Gorghetto:2022sue, Zhang:2023fhs, Chen:2024vgh}. Ground-state vector solitons can be partially or maximally circularly polarized, carrying macroscopic quantum spin~\cite{Zhang:2021xxa, Jain:2021pnk, Zhang:2023ktk}. If dark photons couple to ordinary photons, such polarization could leave imprints signatures on astrophysical electromagnetic signals~\cite{Amin:2023imi}.\footnote{This feature is also expected for spin-2 DM~\cite{Schiappacasse:2025mao}.} The intrinsic polarization of dark photons also offers a unique experimental signature in both astrophysical and terrestrial searches~\cite{Caputo:2021eaa, Amaral:2024tjg, Visinelli:2024wyw, Carenza:2025uwx, Amaral:2024rbj, Amaral:2025fcd}.

Despite their rich phenomenology, the production of ultralight dark photons in the early universe remains a longstanding challenge. A natural possibility is that dark photons are generated via a misalignment mechanism analogous to that proposed for axions. However, reproducing the observed relic abundance through this mechanism typically requires nonminimal couplings to gravity~\cite{Arias:2012az}, which often violate perturbative unitarity or introduce ghost instabilities~\cite{Nakayama:2019rhg, Mou:2022hqb}.\footnote{The vector misalignment mechanism was first discussed in ref.\cite{Nelson:2011sf}, although that work derived the cosmological evolution of a vector field incorrectly. It was later recognized that a nonminimal coupling is required to yield the correct DM relic abundance~\cite{Arias:2012az, Nakayama:2019rhg}.} Under minimal assumptions about their interactions, dark photons can be produced during inflation as isocurvature fluctuations only if their mass satisfies $m_X\gtrsim 10^{-5}\,\mathrm{eV}$~\cite{Graham:2015rva, Kolb:2020fwh, Kolb:2023ydq}. Alternatively, dark photons may be produced through parametric or tachyonic resonance if they couple to an evolving homogeneous scalar field that induces time-dependent effective mass for the dark photon~\cite{Agrawal:2018vin, Co:2018lka, Dror:2018pdh, Adshead:2023qiw, Kitajima:2023pby, Co:2021rhi}. These production mechanisms, however, typically require very large couplings~\cite{Agrawal:2018vin, Co:2018lka, Dror:2018pdh}, entail a degree of fine-tuning~\cite{Adshead:2023qiw}, or neglect the axion self-resonance which could disrupt the desired features~\cite{Kitajima:2023pby, Co:2021rhi}. Dark photons may also be produced via the decay of near-global cosmic strings~\cite{Long:2019lwl}; however, the overall viability of models in which dark photon DM acquires a Higgs mass is challenged by vortex formation~\cite{East:2022rsi, Cyncynates:2023zwj, Cyncynates:2024yxm}.

\begin{figure}
\centering
\begin{minipage}{0.495\linewidth}
\includegraphics[width=\linewidth]{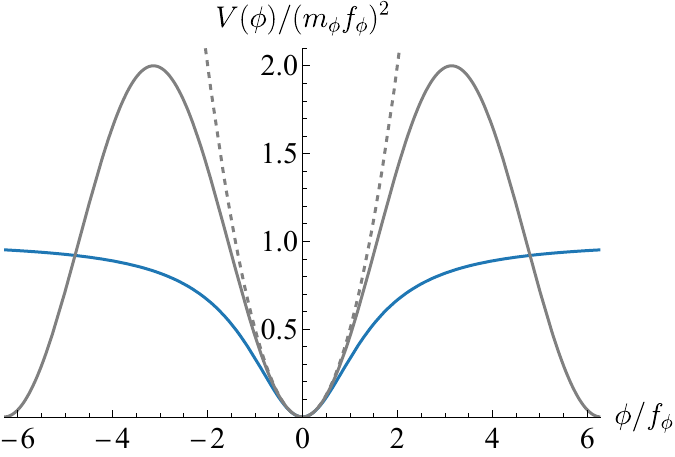}
\end{minipage}
\begin{minipage}{0.42\linewidth}
\includegraphics[width=\linewidth]{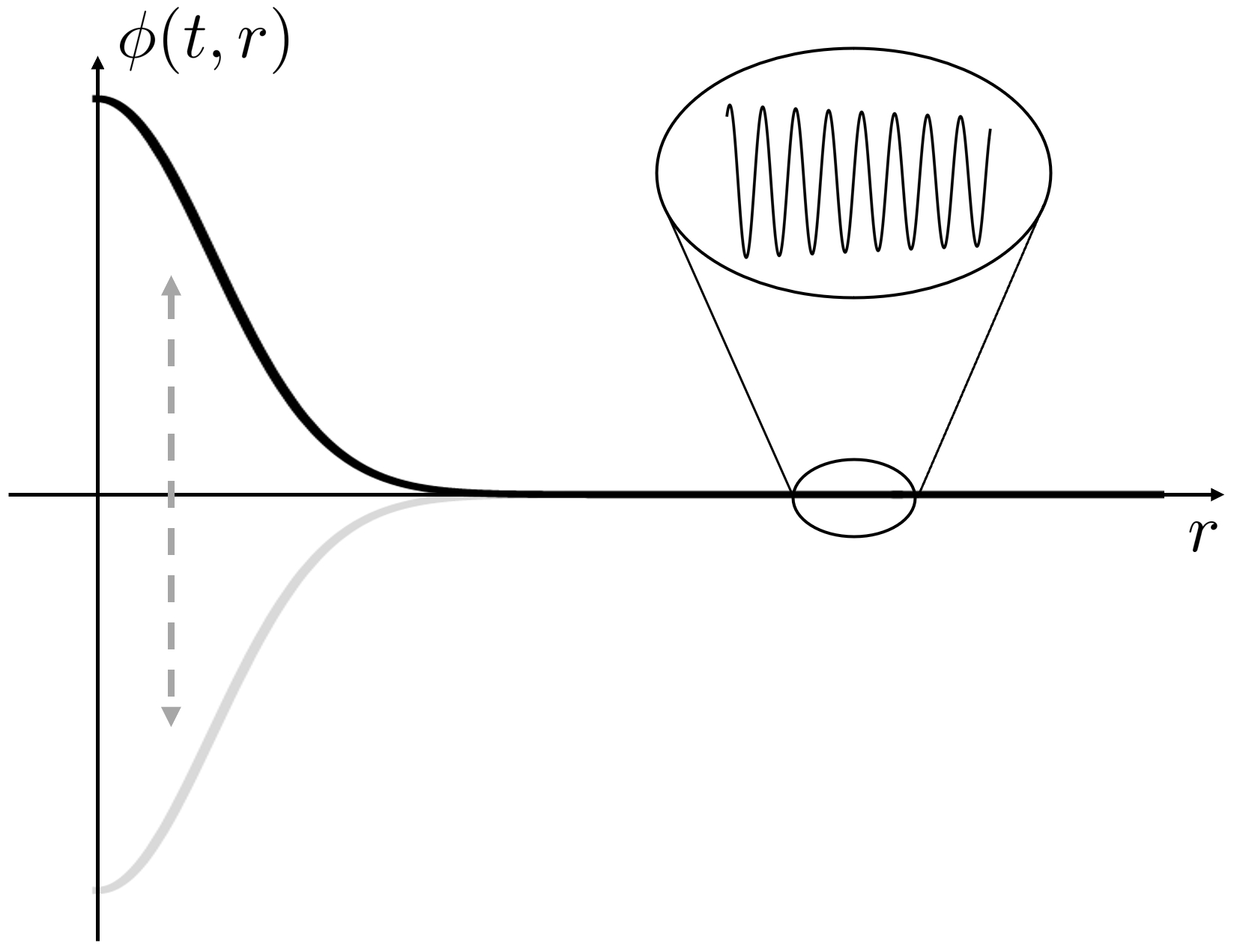}
\end{minipage}
\caption{\textbf{Left:} Comparison between the flattened potential in eq.~\eqref{pot_phi} used in this work (blue), the QCD axion potential $V(\phi) = m_\phi^2 f_\phi^2 [1-\cos(\phi/f_\phi)]$ (solid gray), and a quadratic potential $V(\phi) = m_\phi^2 \phi^2/2$ (dashed gray). \textbf{Right:} Schematic depiction of an oscillon supported by flattened potentials: a longlived, spatially-localized, oscillating field configuration. The field maintains an approximately constant amplitude $\phi_0(t)\sim f_\phi$ within a characteristic radius $R_\mathrm{osc} \simeq \text{a few}\times m_\phi^{-1}$. The oscillon lifetime typically far exceeds $m_\phi^{-1}$ due to suppressed radiation losses.}
\label{fig:oscillon}
\end{figure}

In this work, we propose a new scenario in which efficient dark photon production arises from an oscillating scalar field governed by an unbounded, flattened potential, namely, a potential shallower than quadratic at large field values. For convenience, we loosely refer to this scalar field as an axion, although the mechanism can be generalized to non-axion fields. An example of such a flattened potential is shown in the left panel of figure~\ref{fig:oscillon}. Flattened potentials of this type naturally emerge in multiscalar models~\cite{Dong:2010in, Arvanitaki:2019rax}, string theory constructions~\cite{Witten:1984dg, Svrcek:2006yi, Arvanitaki:2009fg, Arvanitaki:2010sy, Dubovsky:2011tu, McAllister:2014mpa, Kaloper:2016fbr}, and certain Yang-Mills theories~\cite{Nomura:2017ehb, Chatrchyan:2023cmz}. Unlike the QCD axion, which is associated with a periodic potential~\cite{Peccei:1977hh, Weinberg:1977ma, Wilczek:1977pj}, an axion field of this kind can naturally take initial field values well above the effective symmetry-breaking scale $f_\phi$, allowing for significant energy storage in the homogeneous mode.

\begin{figure}
\centering
\includegraphics[width=0.9\linewidth]{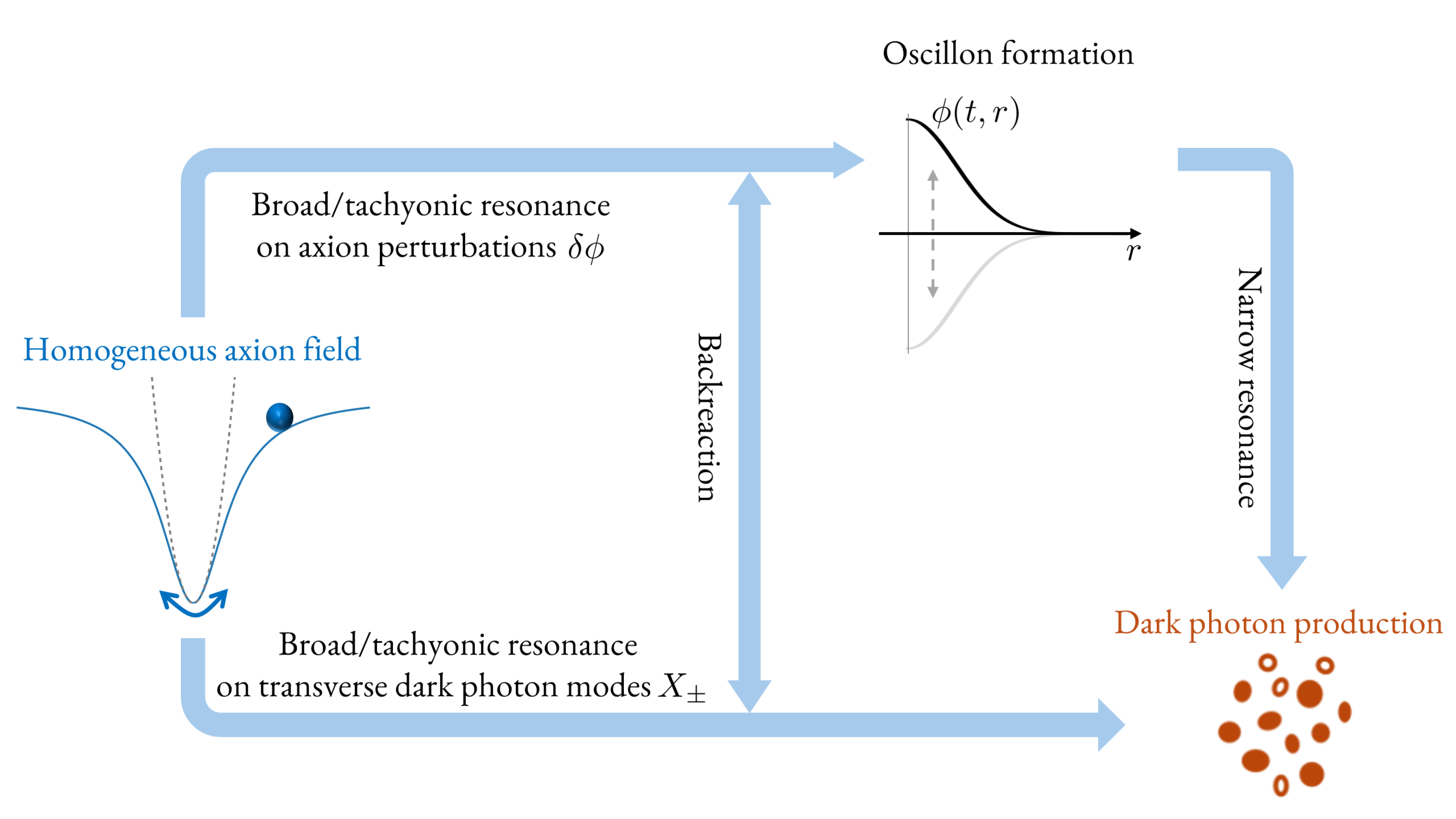}
\caption{Schematic illustration of dark photon production from an oscillating axion field. Initially, the homogeneous axion undergoes parametric resonance,  leading to the exponential growth of axion perturbations and transverse dark photon modes within broad instability bands. The dominant growth channel depends on the axion–dark photon coupling, the initial misalignment angle, and the relative amplitudes of axions and gauge field fluctuations. If axion self-resonance dominates, the system can fragment into localized, longlived oscillons, which could sustain continued dark photon production through parametric resonance in narrower instability bands.}
\label{fig:flowchart}
\end{figure}

Large axion misalignment in flattened potentials gives rise to three key effects. First, the large oscillation amplitude of the axion field around the potential minimum can generate multiple broad instability bands for transverse dark photon modes~\cite{Amin:2014eta}. Second, the delayed onset of axion oscillations reduces the dilution of dark photons caused by Hubble expansion, making the parametric resonance more effective~\cite{Kitajima:2023pby}. Third, flattened potentials naturally facilitate the formation of axion oscillons via self-resonance~\cite{Amin:2011hj, Lozanov:2016hid, Lozanov:2014zfa, Lozanov:2017hjm}, which are localized nonperturbative field configurations with approximately constant amplitudes and long lifetimes~\cite{Copeland:1995fq, Salmi:2012ta, Zhang:2020bec, Zhang:2020ntm, Visinelli:2021uve, Visinelli:2017ooc}. The efficiency of these effects depends on the coupling strength, the relative amplitude of initial inhomogeneous perturbations, and the initial misalignment angles of the axion field. In scenarios where axion self-resonance dominates over resonant dark photon production with broad instability bands, the broad resonance for the dark photon field will be halted due to the developed inhomogeneities. Instead, oscillons can sustain dark photon production as local sources if the axion--dark photon coupling is not too weak, such that the Bose-Einstein statistics could become effective before the produced dark photons leave the objects~\cite{Hertzberg:2018zte, Amin:2020vja}. The overall production process is schematically illustrated in figure~\ref{fig:flowchart}.\footnote{The physical picture discussed here can, in principle, also be realized for QCD axions with the potential $V(\phi) = m_\phi^2 f_\phi^2 [ 1 - \cos(\phi/f_\phi) ]$. However, achieving this scenario requires the axion field to be initially positioned near the top of the potential and leads to a fine-tuning problem~\cite{Arvanitaki:2019rax}.}

In what follows, we first discuss the theoretical and numerical setup and the initial conditions in section \ref{sec:model}. In section \ref{sec:resonance}, we analyze scenarios in which dark photons are produced via broad/tachyonic parametric resonance sourced by a homogeneous oscillating scalar field. In section \ref{sec:oscillon}, we focus on axion self-resonance and demonstrate that dark photons can be produced from localized oscillons. We summarize our results in section \ref{sec:conclusion}. Throughout this work, we adopt the natural units $c=\hbar=1$ and the metric signature $(-,+,+,+)$.

\section{Model and numerical setup}
\label{sec:model}
We consider the following matter action describing an axion–dark photon system:
\begin{align}
    \label{eq:action}
    S_\mathrm{M} = \int d^4x \sqrt{-g} \left[ -\frac{1}{2}\partial_\mu\phi \partial^\mu\phi - V(\phi) - \frac{1}{4} X_{\mu\nu} X^{\mu\nu} - \frac{m_X^2}{2} X_\mu X^\mu - \frac{\alpha}{4f_\phi} \phi X_{\mu\nu} \widetilde X^{\mu\nu} \right] ~,
\end{align}
where $\phi$ is an axion field with potential $V(\phi)$, $X_\mu$ is a dark photon field with mass $m_X$, and $f_\phi$ is an ultraviolet energy scale that quantifies the field excursion of the axion field. The field strength tensor is $X_{\mu\nu} = \partial_\mu X_\nu - \partial_\nu X_\mu$, and its dual is $\widetilde X^{\mu\nu} = \mathcal E^{\mu\nu\rho\sigma} X_{\rho\sigma}/2$, where $\mathcal E^{\mu\nu\rho\sigma}$ is the totally antisymmetric Levi-Civita tensor. The dimensionless parameter $\alpha$ quantifies the coupling strength between axions and dark photons.

We work in a spatially flat, homogeneous, and isotropic Friedmann–Robertson–Walker background, with metric $g_{\mu\nu} = \mathrm{diag} \{-1, a^2(t), a^2(t), a^2(t)\}$ with the scale factor $a(t)$ as a function of the cosmic time $t$. The field equations are given by
\begin{align}
    \label{eq_phi}
    \ddot\phi+ 3H\dot\phi- \frac{\nabla^2}{a^2}\phi+ \partial_\phi V - \frac{\alpha}{a^3 f_\phi} (\nabla X_0 - \dot{\boldsymbol X})\cdot (\nabla\times \boldsymbol X) = 0 ~,\\
    \label{eq_X0}
    \frac{1}{a^2} \nabla\cdot \dot{\boldsymbol X} - \frac{\nabla^2}{a^2} X_0 + m_X^2 X_0 - \frac{\alpha}{a^3 f_\phi} (\nabla\phi) \cdot (\nabla\times\boldsymbol X) = 0 ~,\\
    \label{eq_Xi}
    \ddot {\boldsymbol X} + H \dot{\boldsymbol X} - \frac{\nabla^2}{a^2} \boldsymbol X + 2H\nabla X_0 + m_X^2 \boldsymbol X - \frac{\alpha}{a f_\phi} \left[ \dot\phi\nabla\times \boldsymbol X + (\nabla\phi)\times(\nabla X_0 - \dot{\boldsymbol X}) \right] = 0 ~.
\end{align}
For definiteness, we consider the following axion potential:
\begin{align}
    \label{pot_phi}
    V(\phi) = m_\phi^2\,f_\phi^2\,\frac{\phi^2}{2 f_\phi^2 + \phi^2} ~,
\end{align}
where $m_\phi$ is the mass of the axion field. This potential smoothly interpolates between a quadratic potential in the limit of small oscillations and a nearly flat plateau when $\phi \gg f_\phi$. The potential in eq.~\eqref{pot_phi} can naturally arise from two-scalar theories by integrating out a heavy field~\cite{Dong:2010in, Arvanitaki:2019rax}.

We investigate dark photon production using both linear instability analysis and fully nonlinear lattice simulations. The numerical simulations follow the algorithm developed in ref.~\cite{Agrawal:2018vin}, solving the coupled equations of motion~\eqref{eq_phi}--\eqref{eq_Xi} in a radiation-dominated background, with scale factor $a(t)\propto t^{1/2}$. The simulations are performed in a comoving box of size $L= \pi m_\phi^{-1}$ for results presented in section \ref{sec:resonance} and $L=(\pi/4) m_\phi^{-1}$ in section \ref{sec:oscillon}, both discretized with $128^3$ grid points. The codes are publicly available at \url{https://github.com/hongyi18/AxionDarkPhotonSimulator}. For definiteness, we fix the mass ratio to $m_X/m_\phi = 0.1$ throughout our lattice simulations, while we allow for different mass ratios in our qualitative analysis. Additional discussions regarding different masses ratios are provided in appendix \ref{sec:mass_ratio}. For initial conditions, we place ourselves in the pre-inflationary scenario, thus the axion field prior to the onset of oscillations is taken to be approximately homogeneous, $\phi(\boldsymbol x) = \phi_0 + \delta \phi(\boldsymbol x)$, while the dark photon field is initialized as $X_i(\boldsymbol x) = \delta X_i(\boldsymbol x)$. Here, $\delta\phi$ and $\delta X_i$ are the fluctuations in the axion and dark photon fields, respectively. They are initialized as Gaussian random fields in Fourier space with power spectra given by $P(\boldsymbol k) = c/(2\omega)$, where $\omega=(k^2 + m_{\phi,X}^2)^{1/2}$ is the energy of the mode and $c$ is a tunable constant that parametrizes the amplitude of vacuum fluctuations. By adjusting $c$, we can explore different initial hierarchies between the axion and dark photon field fluctuations. In the following simulations, we will report the resulting magnitude of initial field fluctuations and leave $c$ as an implicit parameter.

Here we briefly comment on the physical mechanisms responsible for generating the initial field fluctuations. If present during inflation, the axion field acquires isocurvature fluctuations with amplitude $\delta\phi \sim H_\mathrm{I}/(2\pi)$, where $H_\mathrm{I}$ is the Hubble parameter during inflation~\cite{Marsh:2015xka, DiLuzio:2020wdo}. The axion field also receives superhorizon-scale fluctuations due to adiabatic perturbations through gravitational interactions, with amplitude $\delta\phi \sim \Phi \phi_0$, where $\Phi\sim 10^{-5}$ denotes the metric perturbation amplitude~\cite{Arvanitaki:2019rax, Planck:2018jri}. Additional axion fluctuations may arise from direct interactions between the axion and the radiation through a temperature-dependent mass~\cite{Kitajima:2021inh, Sikivie:2021trt}. One caveat is that the power spectrum resulting from these mechanisms generally differ from the vacuum spectrum adopted in the simulations; however, this effect is expected to be subdominant since the growth of unstable modes is exponential in time, as will be shown. For the dark photon field, inflation induces isocurvature abundance only in the longitudinal modes, while the transverse modes remain negligible~\cite{Graham:2015rva}. The initial fluctuations in the longitudinal mode are irrelevant for our discussion since they do not undergo exponential growth, as will be demonstrated in the next section. The primary source of fluctuations in the transverse modes is vacuum fluctuations, which follow a Gaussian distribution with a power spectrum $P_{X}(k) = 1/(2\omega_k)$ and are significantly suppressed relative to the axion fluctuations, with $\delta X_i/\phi_0 \sim m_X/(2\pi \phi_0)$.

\section{Dark photon production from broad/tachyonic resonance}
\label{sec:resonance}
The axion field starts to oscillate around the minimum of its potential when $H\simeq H_\mathrm{osc}$, where~\cite{Kitajima:2018zco}
\begin{align}
    \label{eq:condition}
    3H_\mathrm{osc} = \left[\frac{1}{\phi_0}\frac{{\rm d} V(\phi_0)}{{\rm d} \phi_0}\right]^{1/2}\,,
\end{align}
in terms of the initial displacement of the axion field $\phi_0$. For a quadratic potential, the condition is satisfied when $3H \simeq m_\phi$, leading to the onset of coherent oscillations in the axion field at time $t \simeq 1/m_\phi$. For a flattened potential such as the one described in eq.~\eqref{pot_phi} and a large initial field displacement $\phi_0 \gg f_\phi$, the field start to oscillate at much later times than in a quadratic potential since eq.~\eqref{eq:condition} leads to $3H_\mathrm{osc}\ll m_\phi$ and $t_\mathrm{osc} \gg 1/m_\phi$. We argue that this is followed by an efficient dark photon production, as we justify below.

\subsection{Linear instability analysis}
After oscillations begin, the axion field remains homogeneous in the early stages of the evolution. Decomposing the dark photon field in terms of Fourier modes, eqs.~\eqref{eq_X0} and~\eqref{eq_Xi} become
\begin{align}
    \label{eq_X0_k}
    a X_0 + \frac{ik/a}{(k/a)^2 + m_X^2} \dot X_\mathrm{L} = 0 ~,\\
    \label{eq_XL_k}
    \ddot X_\mathrm{L} + \frac{3(k/a)^2 + m_X^2}{(k/a)^2 + m_X^2} H \dot X_\mathrm{L} + [ (k/a)^2 + m_X^2] X_\mathrm{L} = 0 ~,\\
    \label{eq_Xpm_k}
    \ddot X_\pm + H\dot X_\pm + \left( \frac{k^2}{a^2} + m_X^2 \mp \frac{\alpha k}{a f_\phi} \dot\phi\right) X_\pm = 0 ~,
\end{align}
where $X_\mathrm{L}$ and $X_\pm$ are the longitudinal and transverse components of $\boldsymbol X(\boldsymbol k)$. Specifically, we can write $\boldsymbol X(\boldsymbol k) = X_\mathrm{L} \hat{\boldsymbol k} + X_+ \boldsymbol e_+ + X_-\boldsymbol e_-$, where $\hat{\boldsymbol k}$ and $\boldsymbol e_\pm$ are a set of complex orthonormal bases. As we see from eqs.~\eqref{eq_X0_k}--\eqref{eq_Xpm_k}, only the transverse modes $X_\pm$ can be produced from axion oscillations, until the backreaction on axions becomes important and $\phi$ becomes inhomogeneous. During this nonlinear stage, longitudinal modes are expected to form. Furthermore, gravitational interactions could redistribute energy equally between the longitudinal and transverse modes at much later times~\cite{Amaral:2024tjg}.

\begin{figure}
\centering
\includegraphics[width=0.6\linewidth]{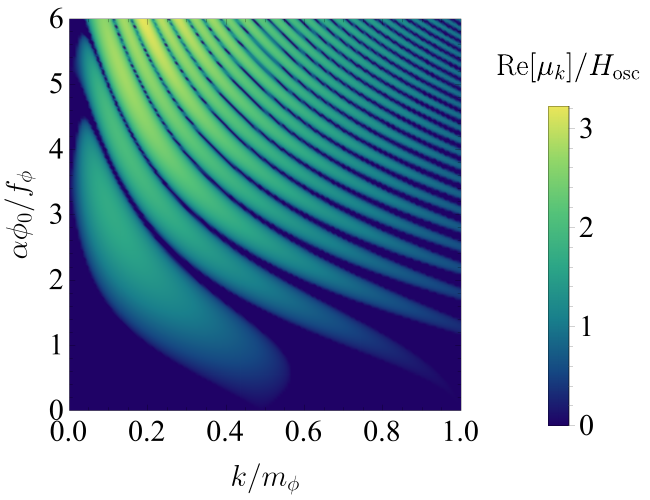}
\caption{Instability diagram for transverse dark photon modes $X_\pm(t,k)$ in flat spacetime, where $\phi_0$ is the oscillation amplitude of the axion field and $H_\mathrm{osc}$ is the Hubble parameter at the onset of oscillations, defined in eq.~\eqref{eq:condition}. The mode $X_\pm(t,k)$ evolves according to eq.~\eqref{floquet_exponent} and can grow exponentially if $\mathrm{Re}[\mu_k]\neq 0$. In an expanding universe, the instantaneous growth rate $\mathrm{Re}[\mu_k]$ competes with the Hubble dilution rate $H(t)$. As the universe expands, the physical wavenumber evolves as $k/a$, the amplitude of axion oscillations decreases as $\phi_0\propto a^{-3/2}$, and the Hubble parameter decreases as $H\propto a^{-2}$ during radiation domination. Here the dark photon mass is set to $m_X=0.1\,m_\phi$.}
\label{fig:instabilitychartvector}
\end{figure}

We can obtain a heuristic understanding of dark photon production by looking at the instability diagram for dark photon perturbations. Neglecting the expansion of the universe and working in the limit where $\phi$ is approximately homogeneous and periodic, Floquet theory allows us to write the evolution of Fourier modes of $X_\pm$ as~\cite{Amin:2014eta}
\begin{align}
\label{floquet_exponent}
X_\pm(t,k) = P_1(t) e^{-\mu_k t} + P_2(t) e^{\mu_k t} ~,
\end{align}
where $P_{1,2}(t)$ are periodic functions and $\mu_k$ is called the Floquet exponent. In particular, if the real part of $\mu_k$ is nonzero and its magnitude is larger than the Hubble parameter $H=1/(2t)$, then the mode could undergo exponential growth. As the scale factor $a(t)$ increases, the physical wave number $k/a$ corresponding a Fourier mode moves through a number of Floquet bands shown in figure~\ref{fig:instabilitychartvector}. Empirically, strong resonance is expected if the maximum value of $\mathrm{Re}[\mu_k]/H$ is greater than $\mathcal O(10)$ as the modes transverse the instability bands~\cite{Amin:2011hj, Amin:2014eta}. Considering that the axion oscillation amplitude decreases as $\phi_0 \propto a^{-3/2}$ and assuming that the Floquet exponent does not change significantly within the instability bands thus $\mathrm{Re}[\mu_k] / H\propto a^2$ during radiation domination, strong resonance can be achieved if $\alpha \phi_0/f_\phi \gtrsim \mathcal{O}(5)$.\footnote{We use $\phi_0$ to denote both the initial field displacement and the amplitude of axion oscillations, which are equivalent in flat spacetime but differ in an expanding universe. Its meaning should be clear based on the context.}

\subsection{Relic abundance of axions and dark photons}
As shown in figure~\ref{fig:instabilitychartvector}, dark photon production via broad or tachyonic resonance is most efficient for modes with physical momentum $k/a \sim 0.2\,m_\phi$. This contrasts with the tree-level expectation, where a single axion decays into two relativistic dark photons, favoring modes with $k/a\approx 0.5\,m_\phi$. When the resonance is efficient and a significant fraction of the axion energy is transferred to dark photons, the resulting axion and dark photon relic abundance can be estimated by
\begin{align}
\label{relic_abundance}
\frac{\rho_\phi}{s} \sim f_\phi\frac{\rho_\phi}{s} \Big|_{H=H_\mathrm{osc}} ~,\quad
\frac{\rho_X}{s} \sim \frac{m_X}{0.2\,m_\phi} f_X\frac{\rho_\phi}{s} \Big|_{H=H_\mathrm{osc}} ~,
\end{align}
where $s=(2\pi^2/45) g_{*S} T^3$ is the entropy density, $f_\phi$ and $f_X$ denote the fractions of the original axion density retained in axions and transferred to dark photons, respectively, and $\rho_\phi|_{H=H_\mathrm{osc}} = b m_\phi^2 f_\phi^2$ is the axion energy density at the onset of oscillations, with $b$ being a model-dependent factor. For the potential in eq.~\eqref{pot_phi}, with the initial axion displacement in the range $5 f_\phi \lesssim \phi_0 < 25 f_\phi$, we find $H_\mathrm{osc}\sim 0.01\,m_\phi$ and $b=1$. The prefactor $m_X/(0.2\,m_\phi)$ in eq.~\eqref{relic_abundance} accounts for the additional redshifting of dark photon energy relative to axions, as the produced dark photons are initially relativistic. The present-day dark photon density parameter is then given by
\begin{align}
\Omega_X h^2 \sim 0.1  f_X b \left( \frac{m_X}{0.1\,m_\phi} \right) \left( \frac{m_\phi}{10^{-17}\,\mathrm{eV}} \right)^{1/2} \left( \frac{f_\phi}{3\times 10^{14}\,\mathrm{GeV}} \right)^{2} \left( \frac{4}{g_*(T_\mathrm{osc})} \right)^{1/4} \left( \frac{0.01 m_\phi}{H_\mathrm{osc}} \right)^{3/2} ~,
\end{align}
where $T_\mathrm{osc}$ is the temperature at $H=H_\mathrm{osc}$, and we have taken $g_*(T_\mathrm{osc}) = g_{*S}(T_\mathrm{osc})$. This expression provides a parametric estimate of the dark photon relic abundance in scenarios where broad or tachyonic resonance dominates the production. As we will show in section~\ref{sec:simulations} from fully nonlinear numerical simulations, the energy density of dark photons typically exceeds that of axions by a factor of $\mathcal{O}(10)$ shortly after the resonance saturates, thus $f_X/f_\phi \sim \mathcal O(10)$. Consequently, dark photons can constitute a dominant component of DM, provided that the mass ratio lies in the range $10^{-3}\lesssim m_X/m_\phi \lesssim 1$. The lower bound ensures that the dark photon relic density exceeds $10\%$ of the total abundance in axions and dark photons, while the upper bound is required for efficient resonance. In particular, the relic abundances of axions and dark photons become comparable when $m_X/m_\phi \sim 10^{-2}$.

The axion field acquires isocurvature fluctuations during inflation with an amplitude $\delta\phi \sim H_\mathrm{I}/(2\pi)$ over the length scale $H_\mathrm{I}^{-1}$. The resulting isocurvature density fluctuation is $\delta_\mathrm{iso} \equiv \delta\rho_\phi/\rho_\phi \sim \delta\phi\partial_\phi b/b$, with $\rho_\phi = b m_\phi^2 f_\phi^2$. This is constrained by CMB observations at the pivot scale $k_0 = 0.05\,\mathrm{Mpc^{-1}}$~\cite{Planck:2018jri}. For a generic power law potential $V(\phi) \propto \phi^n$, one finds $b\propto \phi^n$ and hence $\delta_\mathrm{iso} \sim nH_\mathrm{I} / (2\pi \phi_0)$. If these large-scale isocurvature fluctuations persist through the nonlinear dynamics between axions and dark photons, the CMB bound implies $H_\mathrm{I} \lesssim 6\times 10^{-5} n^{-1} \phi_0$. Depending on the specific model parameters, our scenario (with flattened potentials, $n<2$) can remain viable for both high- and low-scale inflation. In particular, for the potential given in eq.~\eqref{pot_phi} and for a large initial misalignment $\phi_0\gg f_\phi$, the isocurvature constraint can be naturally evaded.

Here we make a comment on the coldness of the dark photon. Since dark photons remain (mildly) relativistic at production, they will free-stream and suppress the formation of small-scale structure in a way analogous to warm DM. We calculate the free-streaming length following ref.~\cite{Garny:2018ali}:
\begin{equation}
    \lambda_\mathrm{fs}= \int_0^{z_{\rm prod}}\frac{v(z)}{H(z)} dz, \label{eq:lambda_fs}
\end{equation}
where $z_\mathrm{prod}$ is the redshift at which dark photons are produced and can be estimated by $H(z_\mathrm{prod})=0.01 m_\phi$. The velocity $v(z)$ is given by 
\begin{equation}
    v(z) = \frac{k_\mathrm{phys}(z)}{E(z)} = \frac{k_\mathrm{phys}(z)}{\sqrt{k_\mathrm{phys}^2(z) + m_X^2}},\quad k_\mathrm{phys}(z) = k_{\rm prod}\frac{1+z}{1+z_{\rm prod}},
\end{equation}
where $k_\mathrm{phys}(z)$ is the physical momentum and $k_{\rm prod} \sim 0.2 m_\phi$ is the peak momentum at production. Taking the mass ratio $m_X/m_\phi=0.1$ and assuming that dark photons constitute all DM, we find that $m_X\gtrsim 10^{-18} \mathrm{eV}$ would be consistent with the Lyman-$\alpha$ constraint on the free-streaming length $\lambda_\mathrm{fs} \lesssim 0.1 ~\mathrm{Mpc}$~\cite{Garny:2018ali,Baur:2017stq}. A more precise treatment of the free-streaming effect would require a full calculation of the matter power spectrum using the full dark photon momentum distribution, which we leave for future work.

\subsection{Simulation results}
\label{sec:simulations}
\begin{figure}
\centering
\begin{minipage}{0.495\linewidth}
\includegraphics[width=\linewidth]{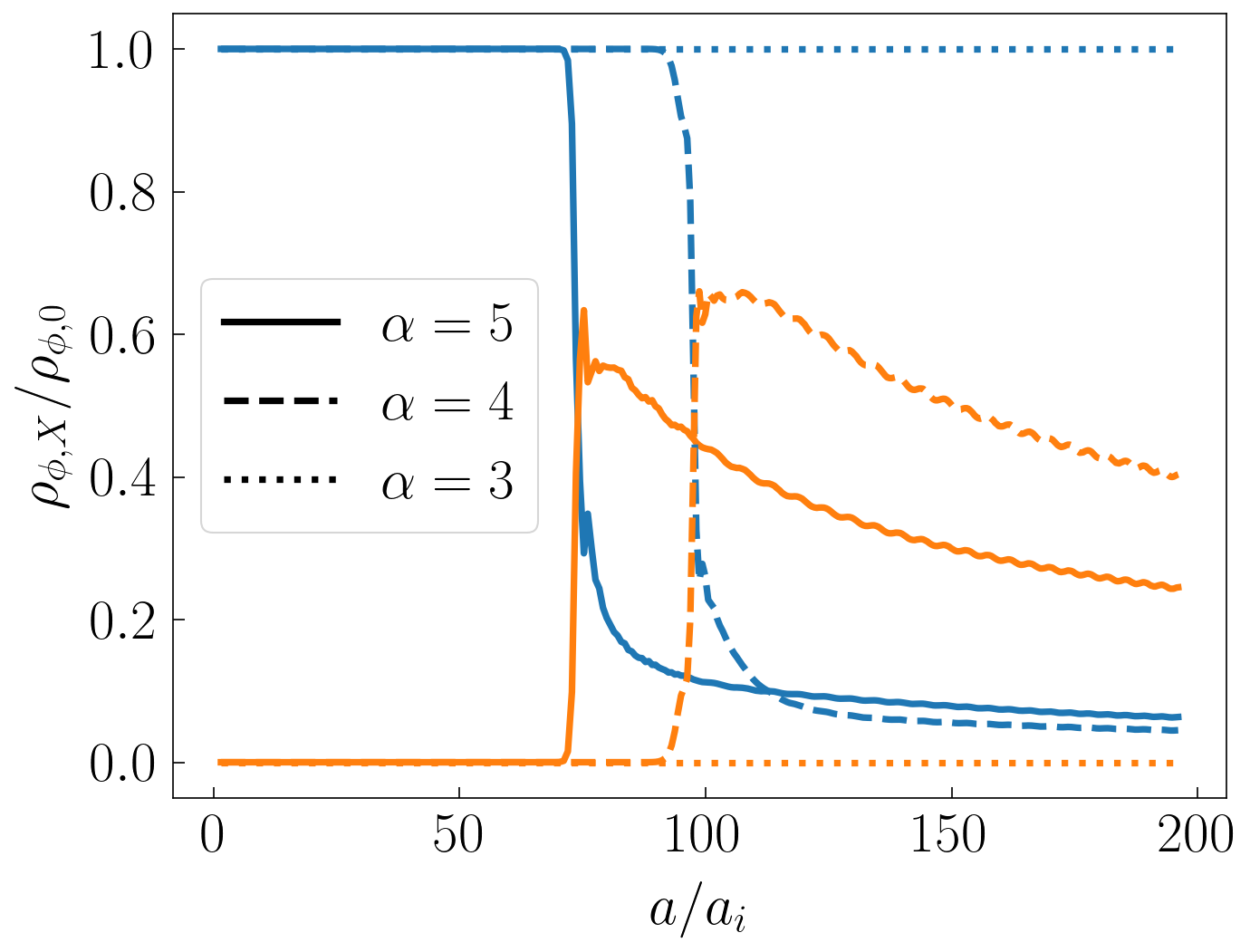}
\end{minipage}
\begin{minipage}{0.495\linewidth}
\includegraphics[width=\linewidth]{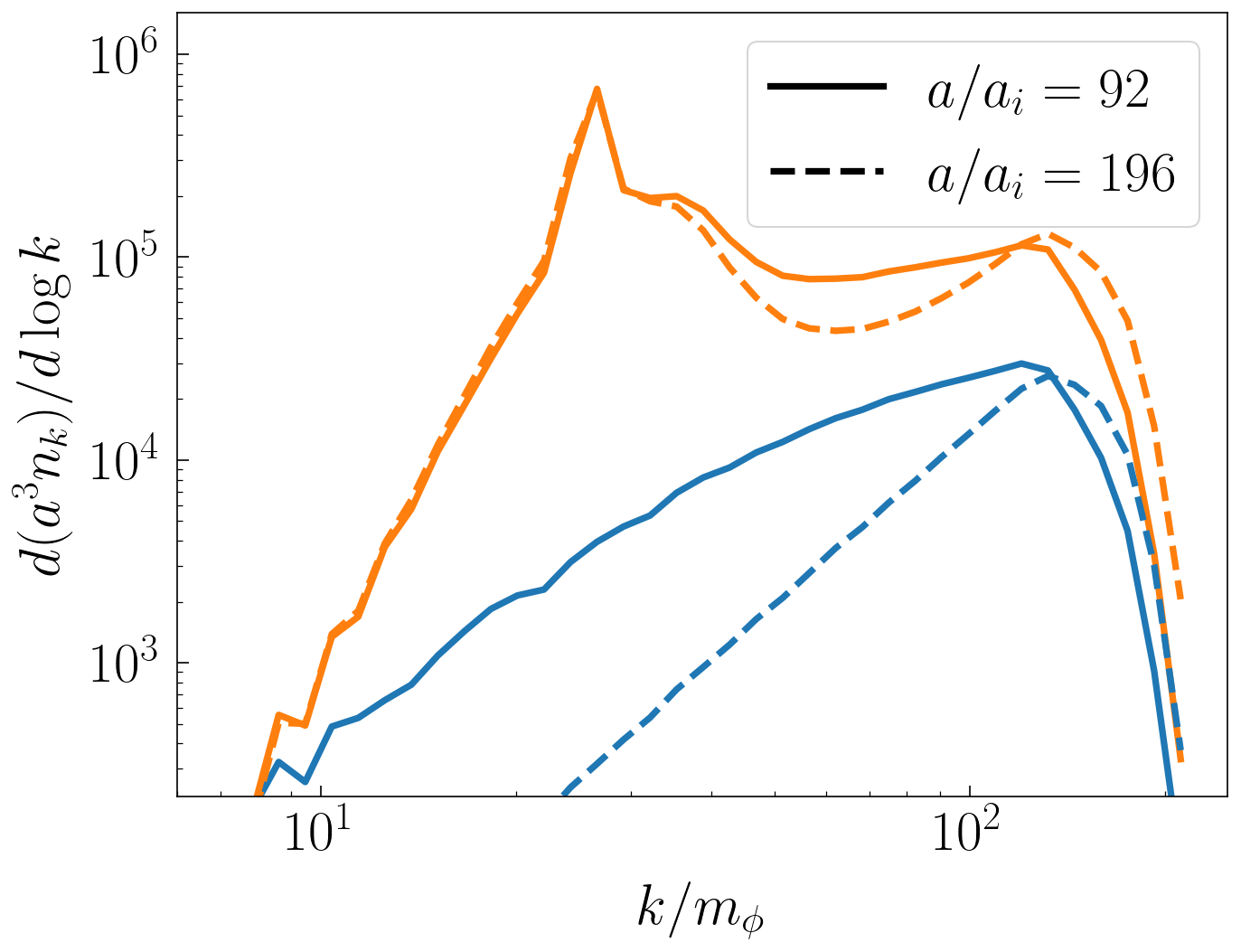}
\end{minipage}
\caption{\textbf{Left:} Evolution of the energy densities of axions (blue) and dark photons (orange) for different values of the coupling constant $\alpha$, normalized to the energy density $\rho_{\phi, 0}$ of a homogeneous axion field with zero coupling. The efficiency of broad resonance decreases for smaller $\alpha$. \textbf{Right:} Spectral number densities of axions (blue) and dark photons (orange) as functions of the comoving wavenumber $k$ for $\alpha=5$ at different times. After the resonance saturates, the dark photon spectrum peaks at physical wave number $k_\mathrm{phys}\sim 0.2\,m_\phi$, corresponding to the instability band for $\alpha\phi_0 \sim f_\phi$ in figure~\ref{fig:instabilitychartvector}. In both panels, the initial conditions are set to $\phi_0=5f_\phi$, $\delta \phi=0$, and $\delta X_i\sim 10^{-35}\phi_0$. The initial scale factor $a_i$ corresponds to the time when $H=10\,m_\phi$.}
\label{fig:noscatinyvecfluc_density_evol}
\end{figure}

In the left panel of figure~\ref{fig:noscatinyvecfluc_density_evol}, we show the evolution of the energy densities of axions and dark photons for different values of the axion–dark photon coupling constant $\alpha$. The efficiency of dark photon production increases with larger values of $\alpha$, given the enhanced strength of parametric resonance. The resonance saturates when the dark photon energy density grows to approximately twice that of axions, signaling that a substantial fraction of the initial axion energy has been effectively transferred to dark photons. In these simulations, the initial field fluctuations are set to $\delta\phi=0$ and $\delta X_i \sim 10^{-35}\,\phi_0$, which are chosen for illustrative purposes.

The right panel of figure~\ref{fig:noscatinyvecfluc_density_evol} shows the spectral number densities of axions and dark photons  at two representative times, $a/a_i=92$ and $a/a_i=196$, fixing $\alpha = 5$. Soon after the production, the dark photon spectrum develops a pronounced peak centered around the physical wavenumber $k_\mathrm{phys} \sim 0.2\, m_\phi$, consistent with the mode that possesses the largest real Floquet exponent in the instability chart of figure~\ref{fig:instabilitychartvector}, for values $\alpha \phi_0 \gtrsim f_\phi$. After the resonance saturates, dark photons continue to interact with axions, leading to the conversion of nonrelativistic axions and mildly relativistic dark photons into those with higher momenta ($k/m_\phi\gtrsim 100$).

\begin{figure}
\centering
\begin{minipage}{0.495\linewidth}
\includegraphics[width=\linewidth]{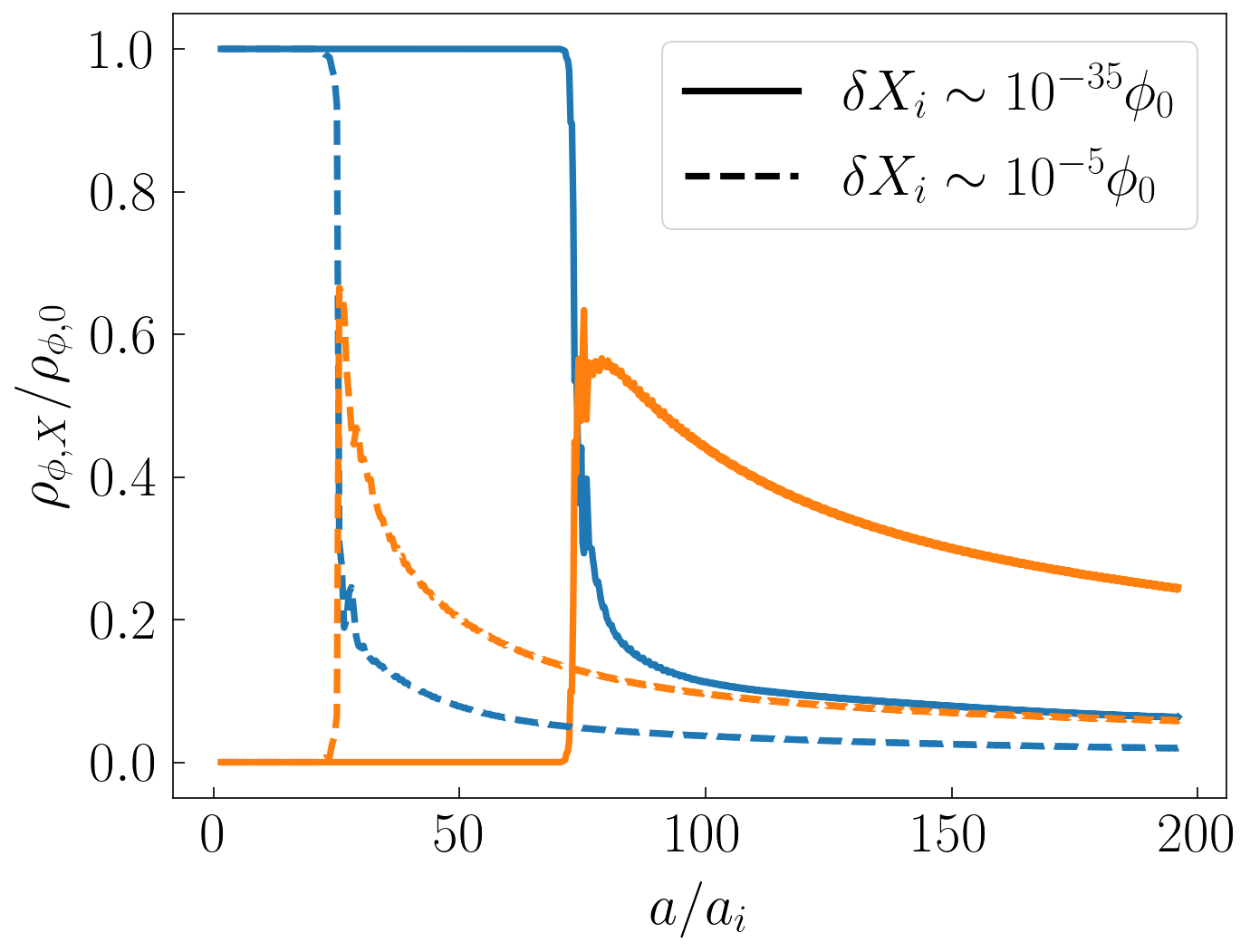}
\end{minipage}
\begin{minipage}{0.495\linewidth}
\includegraphics[width=\linewidth]{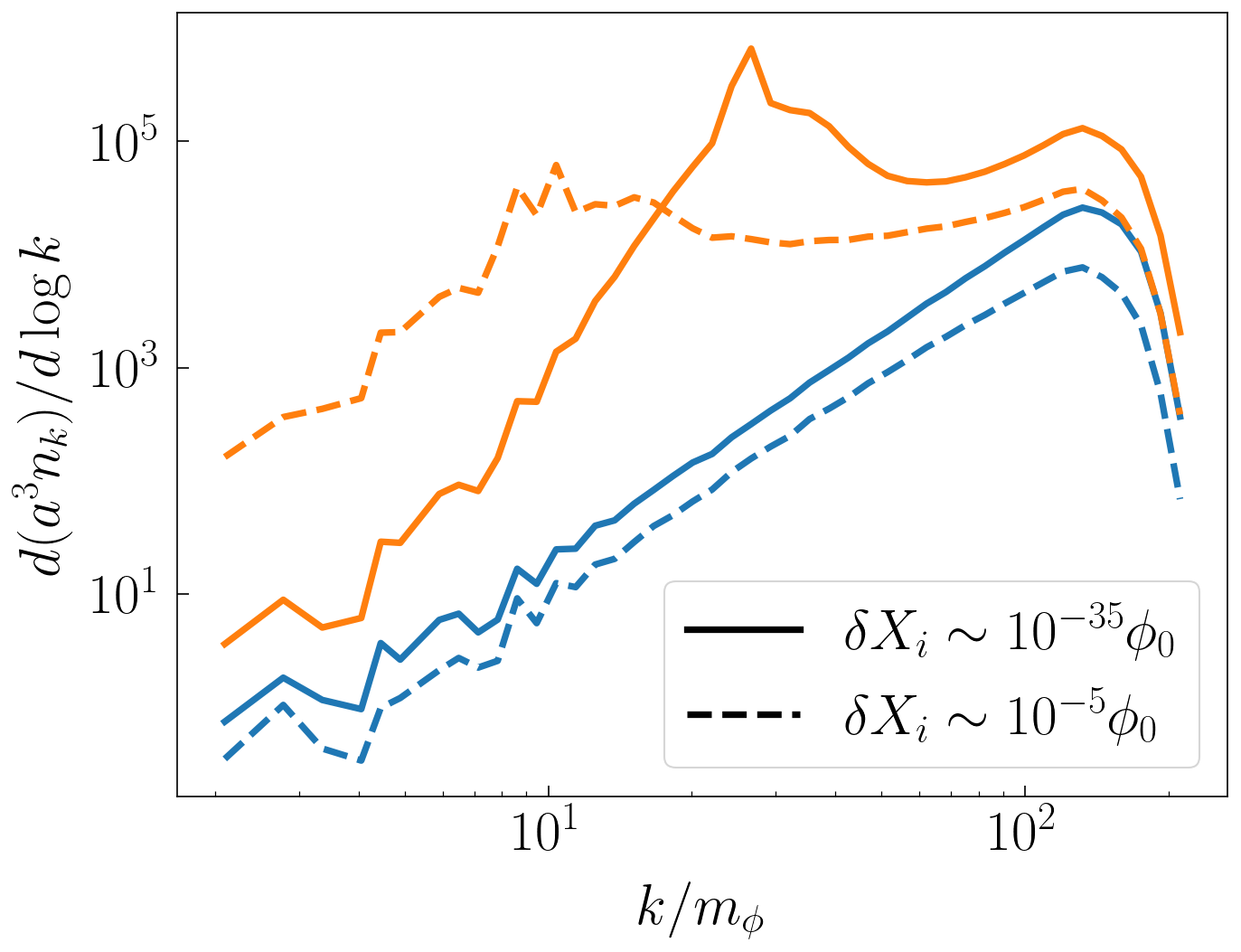}
\end{minipage}
\caption{\textbf{Left:} Evolution of the energy densities of axions (blue) and dark photons (orange) for different initial amplitudes $\delta X_i$ of the dark photon field, using the same normalization as in figure~\ref{fig:noscatinyvecfluc_density_evol}. The efficiency of broad resonance decreases with smaller initial fluctuations in the dark photon field. \textbf{Right:} Spectral number densities of axions (blue) and dark photons (orange) at $a/a_i=196$, for different initial amplitudes of the dark photon field. For larger initial fluctuations $\delta X_i \sim 10^{-5}\,\phi_0$, the resonance is primarily driven by the regime $\alpha \phi_0 \gg f_\phi$, where the instability band remains broad, as shown in figure~\ref{fig:instabilitychartvector}; the resulting dark photon spectrum is less peaked. In both panels, the initial conditions and the coupling constant are set to $\phi_0=5f_\phi$, $\delta \phi=0$, and $\alpha=5$.}
\label{fig:noscafluc_alpha5_density_evol}
\end{figure}

The efficiency of the resonance also depends on the initial amplitude of dark photon fluctuations. In figure~\ref{fig:noscafluc_alpha5_density_evol}, we show the evolution of the energy density for different initial dark photons amplitudes (left) as well as the corresponding spectral number density at the final time of the simulations (right). When the transverse modes of the dark photon field start with a larger amplitude, the resonance saturates earlier. As shown in the right panel, the resulting spectrum for enhanced initial fluctuations exhibits a higher amplitude at low momenta and becomes approximately flat for modes with physical momenta in the range $0.05\,m_\phi \lesssim k_\mathrm{phys} \lesssim 0.5\,m_\phi$. This behavior arises because saturation occurs in the regime where $\alpha \phi_0 \gg f_\phi$, and the corresponding instability band remains broad, as seen in figure~\ref{fig:instabilitychartvector}.

\section{Dark photon production from oscillons}
\label{sec:oscillon}
If the self-resonance of axions becomes important before dark photons are sufficiently produced, which could be the consequence of tiny initial fluctuations of dark photons (as discussed in section \ref{sec:model}) or small couplings between the fields, the inhomogeneity of axion fluctuations $\delta \phi$ would halt the efficient resonant production discussed in previous section. In this case, oscillons could be the main driving force responsible for dark photon production if the coupling constant $\alpha\gtrsim 2$, as we will see shortly.

\subsection{Linear instability analysis}
Oscillons are longlived, spatially-localized, oscillating field configurations supported by attractive self-interactions~\cite{Amin:2011hj, Lozanov:2016hid, Lozanov:2014zfa, Lozanov:2017hjm, Copeland:1995fq, Salmi:2012ta, Zhang:2020bec, Zhang:2020ntm, Visinelli:2021uve, Visinelli:2017ooc}. A schematic plot of an oscillon is shown in figure~\ref{fig:oscillon}. Once formed, oscillons maintain a field amplitude of order $\phi_0(t)\sim f_\phi$ within a characteristic radius $R_\mathrm{osc}\simeq \text{a few}\times  m_\phi^{-1}$, until they eventually decay away.\footnote{In some special cases, oscillons could have a size significantly larger than $m_\phi^{-1}$~\cite{Amin:2010jq}. We do not consider this possibility here.} The approximately constant field amplitudes within oscillons make the resonant production of dark photons occur at later times. To estimate the corresponding growth rate, we neglect both the dark photon mass and the Hubble expansion in eq.~\eqref{eq_Xpm_k} and work in the narrow resonance regime described by $\alpha k/m_\phi \lesssim 1$. The narrow resonance predominantly occurs for modes centered at $k=m_\phi/2$ with bandwidth $\Delta k\simeq \alpha \phi_0/f_\phi$. The associated Floquet exponent for these modes, defined in eq.~\eqref{floquet_exponent}, is approximately $\mu_k\simeq \alpha m_\phi \phi_0/(4 f_\phi)$~\cite{Landau1976Mechanics}. For the Bose-Einstein statistics to be effective, the growth rate $\mu_k$ must exceed the escape rate of dark photons from the oscillon, $(2 R_\mathrm{osc})^{-1}$~\cite{Hertzberg:2018zte}. As a result, axion oscillons can efficiently source dark photon if the coupling satisfies $\alpha \gtrsim \mathcal{O}(1)$. In appendix \ref{sec:coupling}, we discuss how such a coupling can be realized.

\begin{figure}
\centering
\includegraphics[width=0.6\linewidth]{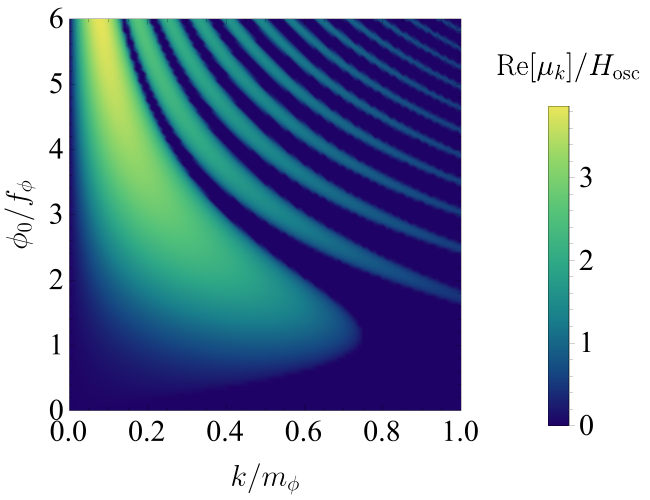}
\caption{Instability diagram for axion perturbations $\delta\phi(t,k)$, with the same notation as in figure \ref{fig:instabilitychartvector}.}
\label{fig:instabilitychartscalar}
\end{figure}

As attractor solutions, oscillons could emerge due to self-resonance of axion perturbations. Decomposing the axion field into $\phi(t,\boldsymbol x) = \bar\phi(t) + \delta\phi(t,\boldsymbol x)$ and neglecting the backreaction of dark photons, the linearized equation governing axion perturbations is given by
\begin{align}
\ddot{\delta\phi} + 3H\dot{\delta\phi} + \left[ \frac{k^2}{a^2} + \partial_{\bar\phi}^2 V(\bar\phi) \right] \delta\phi = 0 ~.
\end{align}
For a (quasi-)periodic background $\bar\phi(t)$, axion perturbations $\delta\phi$ can experience self-resonance and grow exponentially. The instability diagram in flat spacetime is shown in figure~\ref{fig:instabilitychartscalar}, where several broad resonance bands appear. These bands enable the efficient growth of inhomogeneities and facilitate the formation of oscillons.\footnote{The detailed formation process (including how perturbations backreact on the homogeneoues background and distribute into oscillons) is irrelevant to our disucssions since oscillons are attractor solutions and intermediate processes do not affect their final configurations significantly. For interested readers, see~\cite{Copeland:1995fq, Salmi:2012ta, Zhang:2020bec}.}
Achieving strong resonance requires field amplitudes satisfying $\phi_0/f_\phi\gtrsim \mathcal{O}(5)$.

\subsection{Simulation results}
\begin{figure}
\centering
\begin{minipage}{0.495\linewidth}
\includegraphics[width=\linewidth]{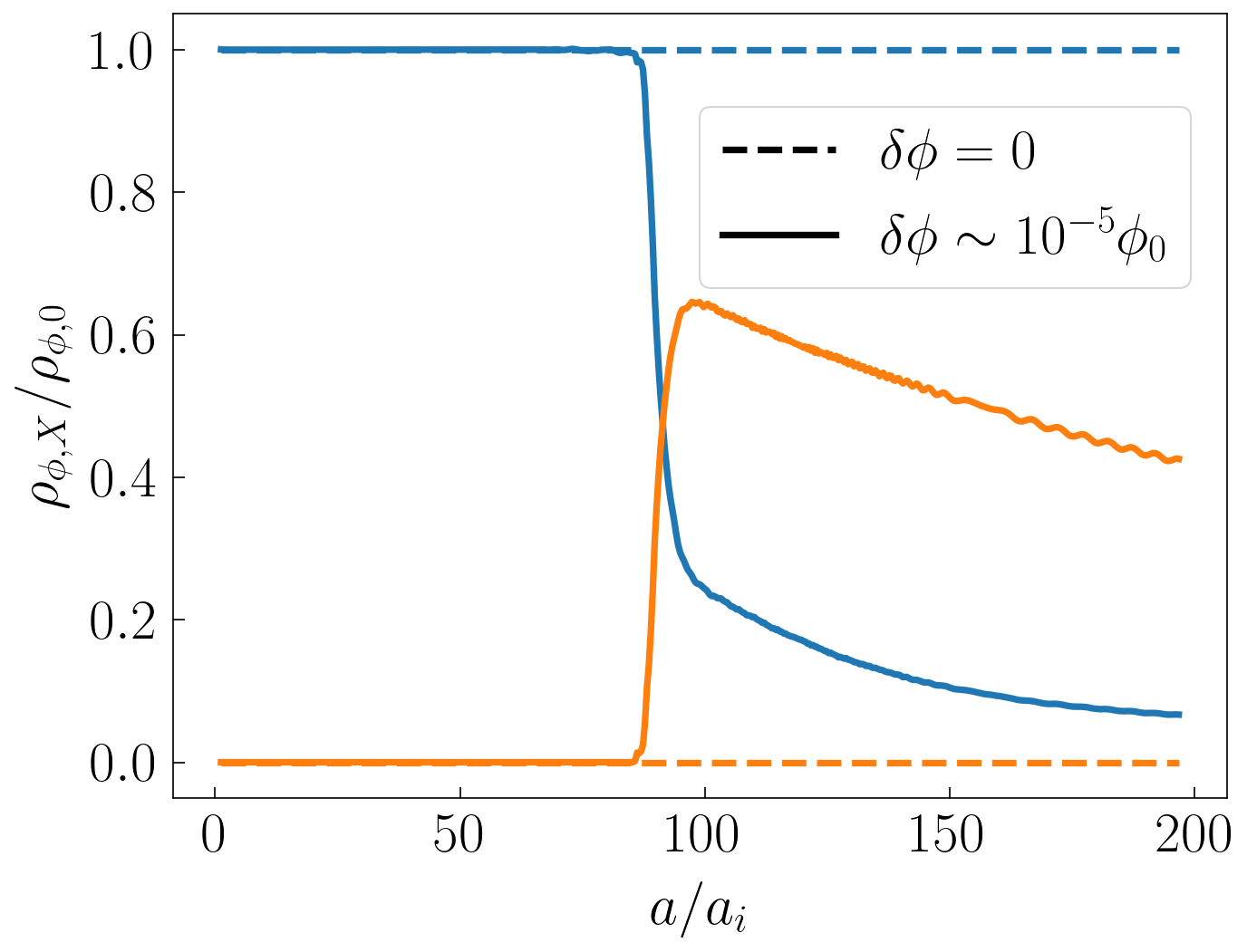}
\end{minipage}
\begin{minipage}{0.495\linewidth}
\includegraphics[width=\linewidth]{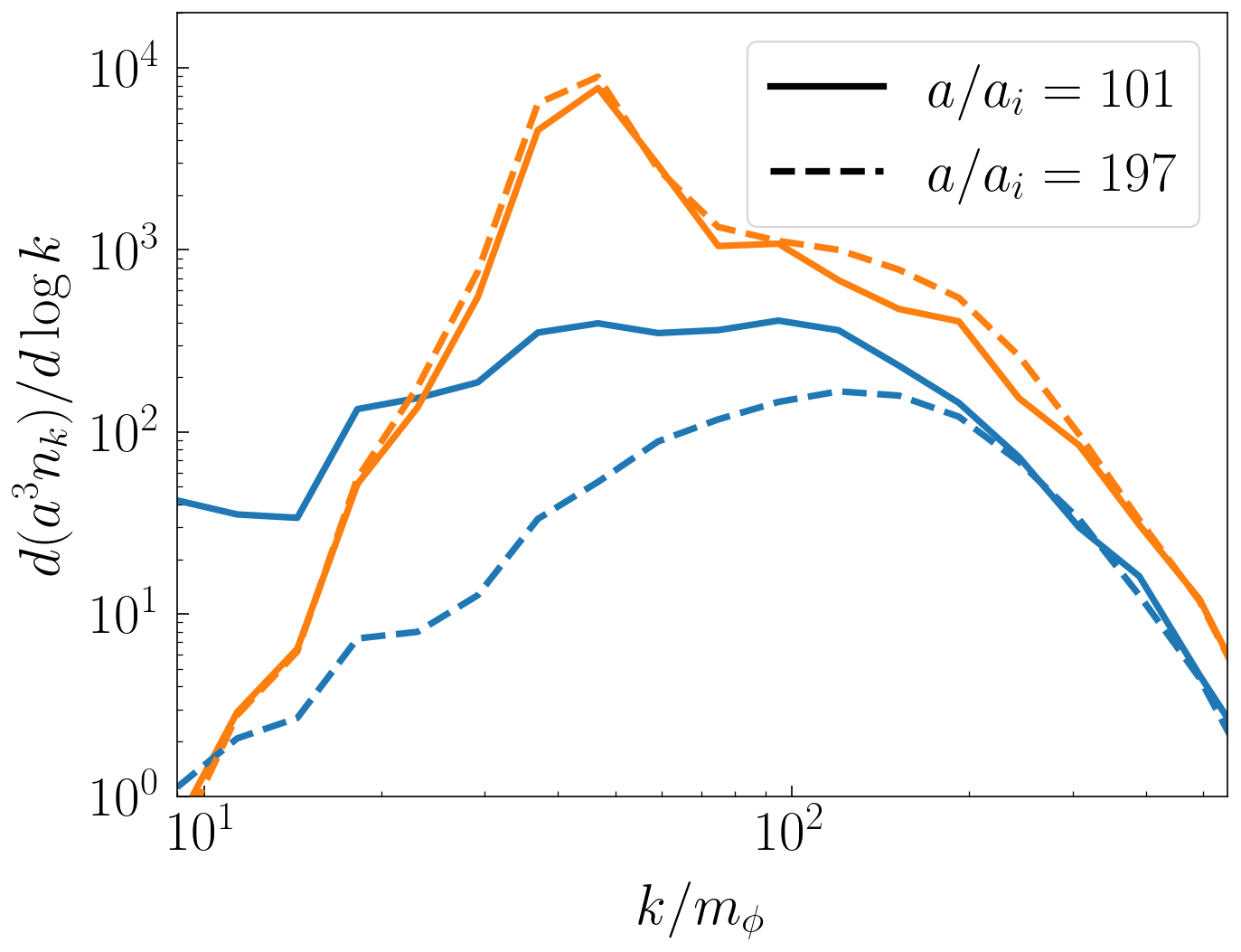}
\end{minipage}
\caption{\textbf{Left:} Evolution of the energy densities of axions (blue) and dark photons (orange) for different initial amplitudes of the axion field fluctuations, normalized as in figure~\ref{fig:noscatinyvecfluc_density_evol}. Axion self-resonance becomes efficient for larger initial fluctuations, $\delta\phi \sim 10^{-5}\phi_0$, leading to the formation of oscillons, which act as localized sources for dark photon production. \textbf{Right:} Spectral number densities of axions (blue) and dark photons (orange) for $\delta\phi\sim 10^{-5}\phi_0$ at different times. Dark photon production within oscillons proceeds via narrow parametric resonance, resulting in a spectrum that peaks at the physical wavenumber $k_\mathrm{phys} \simeq 0.5\,m_\phi$. In both panels, the initial conditions and the coupling constant are set to $\phi_0 = 5\,f_\phi$, $\delta X_i\sim 10^{-35}\,\phi_0$, and $\alpha=2$.}
\label{fig:tinyvecfluc_alpha2_density_evol}
\end{figure}

In the left panel of figure~\ref{fig:tinyvecfluc_alpha2_density_evol}, we show the evolution of the energy densities of axions and dark photons for different initial amplitudes of the axion field fluctuations. As indicated by the dashed lines corresponding to $\delta\phi=0$, dark photon production via broad or tachyonic resonance is inefficient if the coupling and the initial axion misalignment are not sufficiently large; see also the $\alpha=3$ curve in the left panel of figure \ref{fig:noscatinyvecfluc_density_evol}. However, when the initial axion fluctuations are larger (solid curves), efficient dark photon production becomes possible again. This enhancement is driven by the formation of oscillons, which act as localized, longlived sources that stimulate dark photon production.

The right panel of figure~\ref{fig:tinyvecfluc_alpha2_density_evol} displays the spectral number densities of axions and dark photons at $a/a_i = 101$ and $a/a_i=197$. The spectrum of dark photons is sharply peaked at the physical momentum $k/a \simeq 0.5\,m_\phi$, consistent with narrow parametric resonance occurring within axion oscillons. After the resonance saturates, energy continues to transfer from low-momentum axions into higher-momentum dark photons.

\begin{figure}
\centering
\begin{minipage}{0.57\linewidth}
\includegraphics[width=\linewidth]{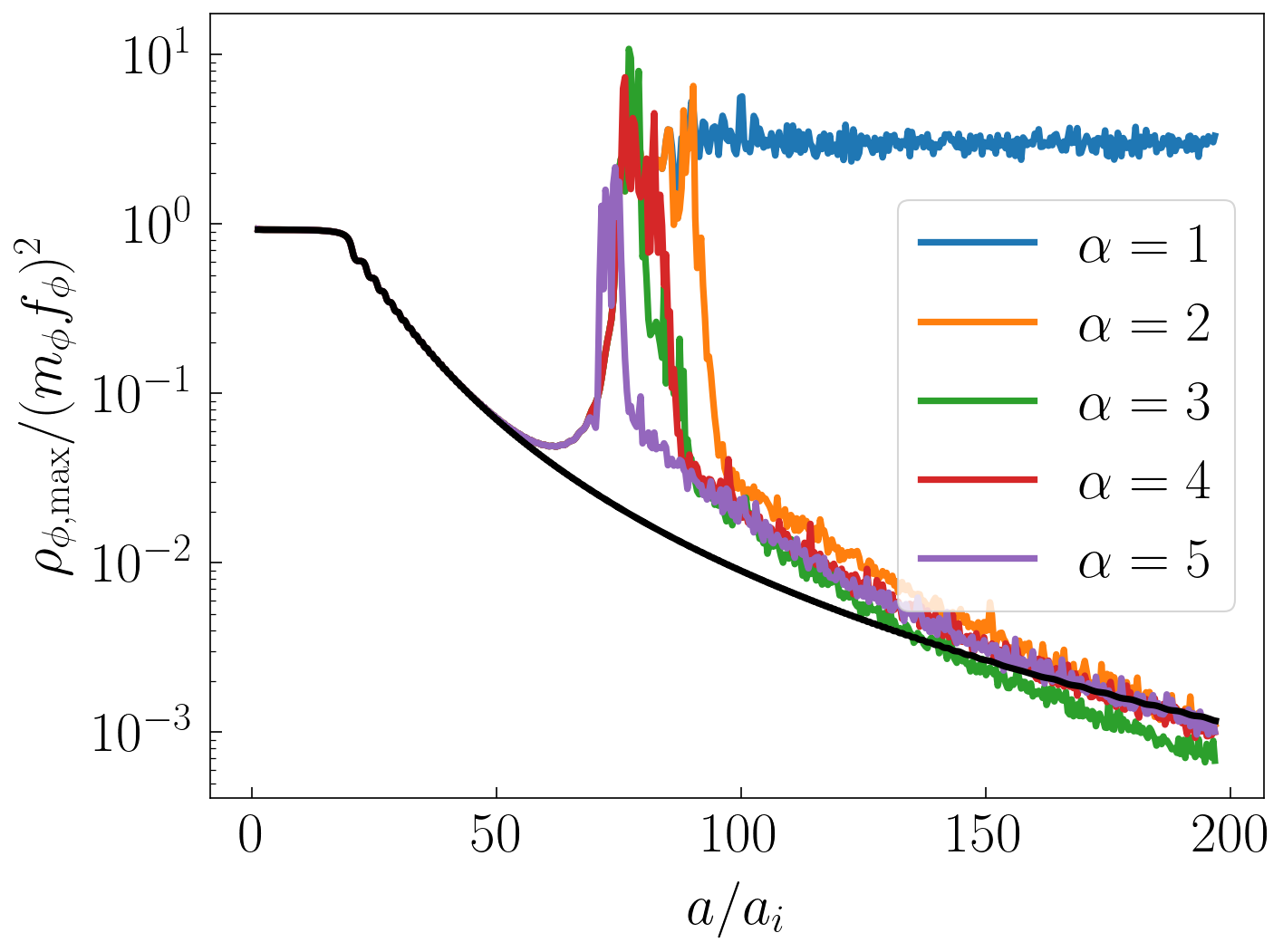}
\end{minipage}
\begin{minipage}{0.42\linewidth}
\includegraphics[width=\linewidth]{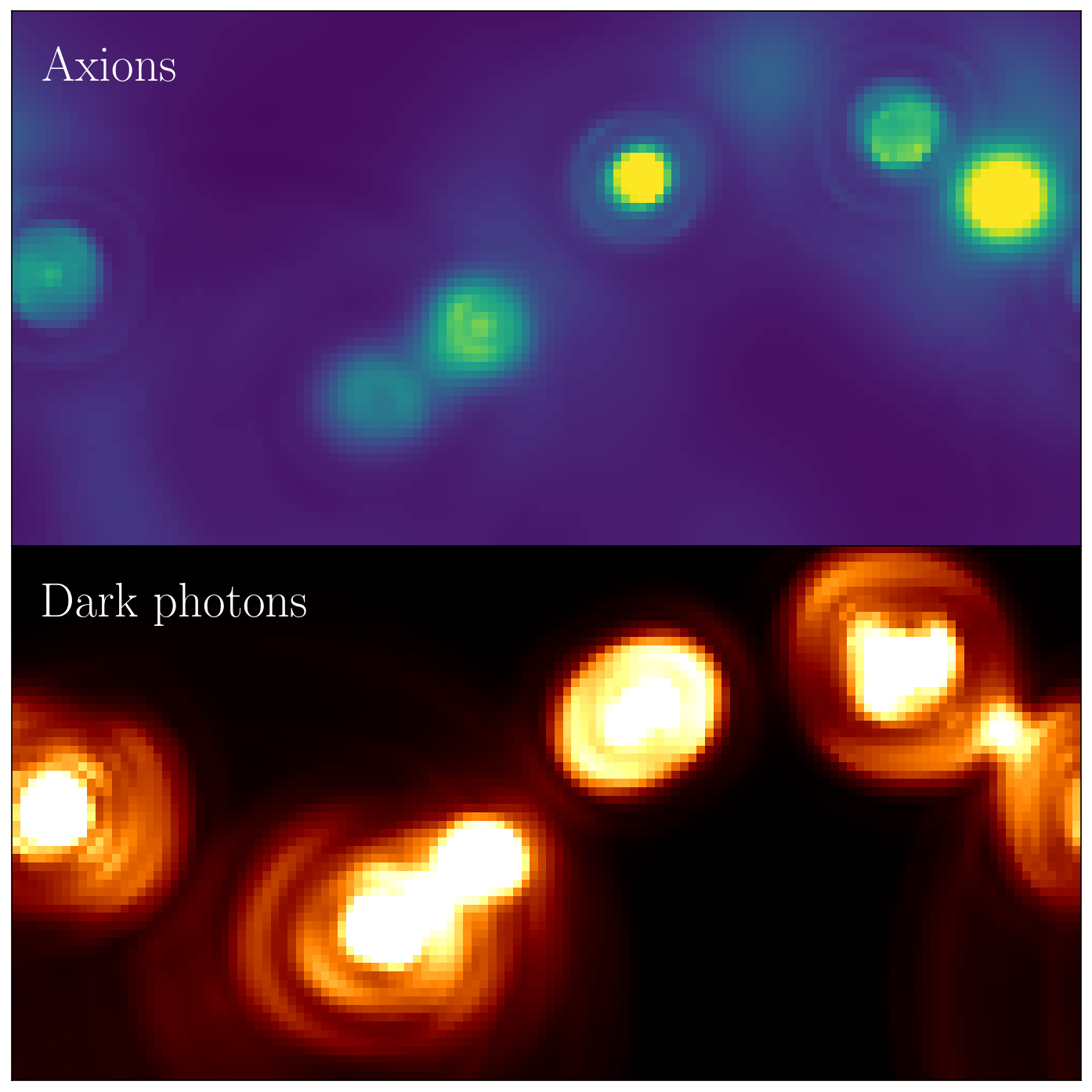}
\end{minipage}
\caption{\textbf{Left:} Evolution of the maximum axion density $\rho_{\phi, \mathrm{max}}$ for different values of the coupling constant $\alpha$. The increase of $\rho_{\phi, \mathrm{max}}$, driven by axion self-resonance, facilitates the formation of oscillons, within which the maximum density remains approximately constant. These oscillons can subsequently convert into dark photons, provided $\alpha\gtrsim 2$. For comparison, the case of a homogeneous axion field with zero coupling is shown as the black solid line. Here the initial conditions are set to $\phi_0=5f_\phi$, $\delta \phi\sim 10^{-5}\,\phi_0$, and $\delta X_i\sim 10^{-35}\,\phi_0$. \textbf{Right:} Snapshot of the energy densities of axions and dark photons at $a/a_i=88.8$ for $\alpha=2$.}
\label{fig:tinyvecflucrhomaxphi}
\end{figure}

In the left panel of figure~\ref{fig:tinyvecflucrhomaxphi}, we present the evolution of the maximum energy density of the axion field for various values of the coupling constant $\alpha$. The maximum density begins to increase around $a/a_i\simeq 60$, followed by a rapid decline. For moderate couplings, e.g., $\alpha=2$, the maximum density plateaus and remains nearly constant over an extended period, indicating the formation of oscillons. Furthermore, we find that efficient energy transfer from oscillons to dark photons requires a coupling strength $\alpha\gtrsim 2$, in agreement with the expectations from our linear instability analysis. In the right panel, we present the energy densities of axions and dark photons at $a/a_i=88.8$ for $\alpha=2$, which illustrates the role of oscillons as local sources for dark photon production.

\section{Discussion and conclusions}
\label{sec:conclusion}

We have proposed a novel and efficient mechanism for the production of ultralight dark photon DM in the early universe, driven by an oscillating scalar field with a flattened, unbounded potential. Such potentials naturally arise in multiscalar extensions of field theory and in various string theory constructions, and they allow for initial field displacement significantly larger than the characteristic mass scale $f_\phi$. Our framework relies on a moderately large initial misalignment of the homogeneous scalar field, which facilitates dark photon production through parametric resonance. This enhancement results from both the large amplitude of oscillations and the delayed onset of scalar field oscillations. We have shown that, depending on the interplay between the axion–dark photon coupling and the role of axion self-resonance, dark photons can be efficiently produced via either broad/tachyonic resonance from the homogeneous oscillating background field, or narrow resonance localized within oscillons. As demonstrated by our lattice simulations, the latter channel becomes particularly effective when the coupling satisfies $\alpha\gtrsim 2$.

In this setup, dark photons can account for a substantial fraction of the total DM abundance, amounting to up to $\gtrsim \mathcal O(10)\%$, for a wide range of mass ratios, $m_X/m_\phi \sim \mathcal O (10^{-3}$--$1)$. Our result provides a viable and less fine-tuned mechanism for generating ultralight dark photon DM with masses $m_X \gtrsim 10^{-18} \mathrm{eV}$ as the dominant DM component, while remaining consistent with existing cosmological and astrophysical constraints.

Our scenario remains valid under different hierarchies of initial fluctuations between the axion and the transverse components of the dark photon field, denoted $\delta\phi$ and $\delta X_\pm$. The most natural expectation is $\delta\phi \gg \delta X_\pm$ since the axion acquires relatively large curvature fluctuations from its homogeneous background. In this regime, axion self-resonance is highly efficient unless the coupling is sufficiently large, and dark photon production can persist within oscillons. Additional sources of field fluctuations include, for example, inflationary fluctuations (for axions only) and vacuum fluctuations. Depending on the hierarchy, dark photon production may proceed through either channel discussed earlier. See section \ref{sec:model} for more details on initial conditions.

From the point of view of ultraviolet completions, achieving an effective axion-dark photon coupling of order unity in eq.~\eqref{eq:action} can be done naturally in several models. As we discuss in appendix \ref{sec:coupling}, a dimensionless parametrization of the coupling in a U(1) dark sector includes an anomaly coefficient $C_\mathrm{D}$, a dark fine-structure constant $\alpha_\mathrm{D}$, and an enhancement factor $j$ coming from additional dynamics, such as the clockwork mechanism \cite{Agrawal:2018vin, Choi:2015fiu, Higaki:2016yqk} or mixing with another scalar. New states introduced in these ultraviolet completions should align with collider, cosmological or perturbative consistency requirements.

In this work we do not assume any particular origin of the dark photon mass (e.g., Higgs, Stueckelberg, or Proca), since dark photon production in our scenario relies only on the presence of a nonzero mass term for the vector field. However, if the mass arises through a Higgs mechanism, additional effects may become relevant. In particular, the presence of a dark Higgs field opens the possibility of topological defect formation, such as vortices or cosmic strings, which can affect the abundance, phenomenology, and stability of dark photon DM. In fact, vortex formation has been argued to jeopardize the viability of Higgsed dark photon DM \cite{East:2022rsi, Cyncynates:2023zwj, Cyncynates:2024yxm}, making alternative realizations such as the one explored here especially compelling. While our results on axion-induced resonance remain valid in the Higgs case, a full treatment would require incorporating the additional dynamics of the dark Higgs sector. Future work may investigate this scenario and the associated observable signatures, such as gravitational wave production, impacts on small-scale structure formation, and potential signals in direct or indirect DM detection experiments. These investigations will offer new insights for probing the dark sector, and provide with a new channel for the production of ultralight vector fields.

\acknowledgments
We would especially like to thank Mustafa A.\ Amin for useful discussions. We would also like to thank Andrew J.\ Long and Sudhakantha Girmohanta for helpful comments. HYZ and LV acknowledge support by the National Natural Science Foundation of China (NSFC) through the grant No.~12350610240 ``Astrophysical Axion Laboratories''. LV also thanks Istituto Nazionale di Fisica Nucleare (INFN) through the ``QGSKY'' Iniziativa Specifica project. AC is supported by the National Natural Science Foundation of China (NSFC) through the grant No.~12090064. EDS acknowledges support from the FONDECYT project No.\ 1251141 (Agencia Nacional de Investigaci\'on y Desarrollo, Chile). PA acknowledges support from FONDECYT project No.~1251613. LR is supported by Teaming for Excellence grant Astrocent Plus (GA: 101137080) funded by the European Union and the grant ‘AstroCeNT: Particle Astrophysics Science and Technology Centre’ carried out within the International Research Agendas programme of the Foundation for Polish Science financed by the European Union under the European Regional Development Fund. This publication is based upon work from the COST Actions ``COSMIC WISPers'' (CA21106) and ``Addressing observational tensions in cosmology with systematics and fundamental physics (CosmoVerse)'' (CA21136), both supported by COST (European Cooperation in Science and Technology).

\appendix

\section{Floquet analyses for different mass ratios}
\label{sec:mass_ratio}

\begin{figure}
\centering
\begin{minipage}{0.495\linewidth}
\includegraphics[width=\linewidth]{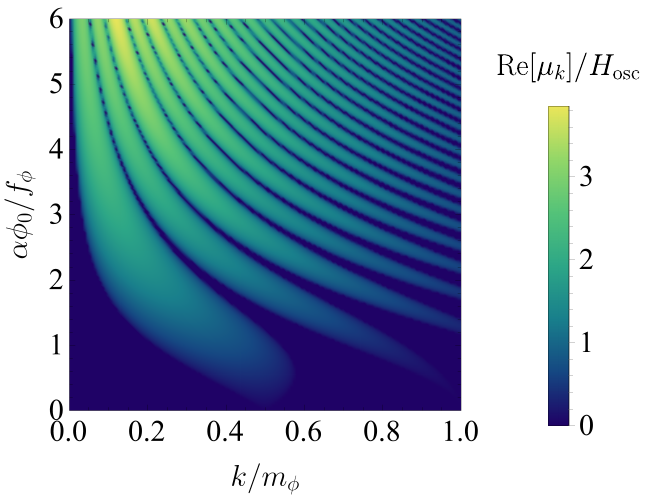}
\end{minipage}
\begin{minipage}{0.495\linewidth}
\includegraphics[width=\linewidth]{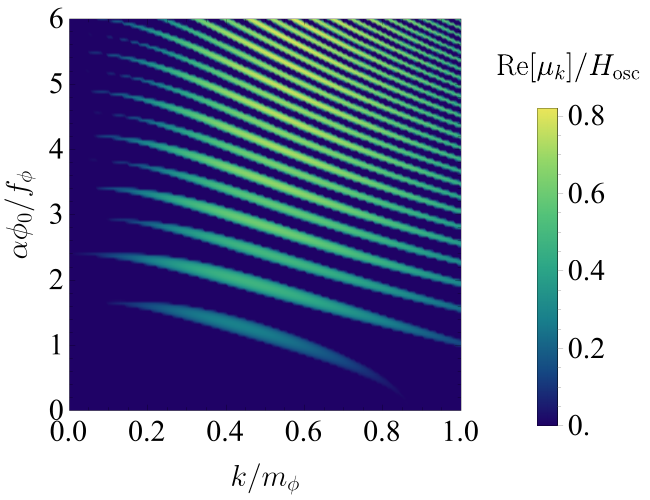}
\end{minipage}
\caption{Instability diagram for transverse dark photon modes $X_\pm(t,k)$ for massless dark photons (left) and the mass ratio $m_X/m_\phi=0.5$ (right), with the same notation as in figure~\ref{fig:instabilitychartvector}.}
\label{fig:instabilitychartvectormX2mphi0}
\end{figure}

In figure~\ref{fig:instabilitychartvectormX2mphi0}, we present additional Floquet analyses for the cases of massless dark photons and a mass ratio $m_X/m_\phi = 0.5$. Compared to the case with $m_X/m_\phi=0.1$ shown in figure~\ref{fig:instabilitychartvector} in the main text, the instability region extends to lower momenta $k\lesssim 0.01 m_\phi$ for massless dark photons, while it becomes narrower when the dark photon mass equals half of the axion mass. This behavior can be understood from eq.~\eqref{eq_Xpm_k}, which shows that tachyonic instability occurs only for modes satisfying
\begin{align}
\frac{\alpha \phi_0}{2f_\phi} - \sqrt{\frac{\alpha^2 \phi_0^2}{4 f_\phi^2}-\frac{m_X^2}{m_\phi^2}} \lesssim \frac{k}{a m_\phi} \lesssim \frac{\alpha \phi_0}{2f_\phi} + \sqrt{\frac{\alpha^2 \phi_0^2}{4 f_\phi^2}-\frac{m_X^2}{m_\phi^2}} ~,
\end{align}
assuming $\dot\phi \sim m_\phi \phi_0$. Therefore, as the dark photon mass increases, the instability band narrows, thereby reducing the efficiency of tachyonic resonance. Conversely, as the dark photon mass decreases, the lower bound of the instability band shifts to smaller momenta, enabling tachyonic growth at low $k$. Moreover, a larger axion–dark photon coupling $\alpha$ or a larger initial misalignment $\phi_0$ broadens the instability band, thereby making efficient resonance possible even for heavier dark photon masses.

\section{Axion-dark photon coupling}
\label{sec:coupling}

As demonstrated in the main text, efficient broad or tachyonic resonance requires $\alpha \phi_0 \gg f_\phi$ (a condition that may also be achieved through kinetic misalignment) and successful dark photon production from oscillons requires $\alpha\gtrsim 2$. If the initial misalignment is only moderately large (e.g., $\phi_0 \lesssim 10 f_\phi$), then realizing our mechanism requires $\alpha \gtrsim \mathcal O(1)$.

In the context of a $U(1)$ gauge theory and a natural parametrization of the coupling takes the form
\begin{align}
    \label{eq:alpha}
    \alpha = j \frac{C_\mathrm{D}\,\alpha_\mathrm{D}}{2\pi}~,
\end{align}
where $C_\mathrm{D}$ is the anomaly coefficient, which can be large by introducing fermions with large charges or by increasing the number of charged fermion species, $\alpha_\mathrm{D}$ denotes the dark fine-structure constant, and $j$ is an additional model-dependent enhancement factor. The requirement of perturbativity imposes the constraint $C_\mathrm{D}\alpha_\mathrm{D}<1$, so achieving $\alpha\gtrsim \mathcal O(1)$ generally necessitates a large enhancement factor, $j\gg 1$. In the following we discuss two ways to obtain a large $j$ factor.

One way to achieve a significant axion--dark photon coupling is to incorporate multiple symmetry-breaking scalar fields, similar to what is done in the clockwork mechanism~\cite{Agrawal:2018vin, Choi:2015fiu, Higaki:2016yqk}. Another route to enhancement involves mixing between two scalar fields, where only one directly couples to dark photons. In the case of kinetic mixing with coupling $\epsilon$ or mass mixing through a term $m_{12}^2 \phi_1 \phi_2$, diagonalization of the kinetic or mass matrices induces an effective coupling of the second scalar field to dark photons. The resulting enhancement factor $j$ is of order $\epsilon$ or $m_{12}^2/(m_2^2 - m_1^2)$.

\bibliographystyle{jhep}
\bibliography{ref}

\providecommand{\href}[2]{#2}\begingroup\raggedright\begin{thebibliography}{10}

\bibitem{Bertone:2004pz}
G.~Bertone, D.~Hooper and J.~Silk, \emph{{Particle dark matter: Evidence,
  candidates and constraints}},
  \href{https://doi.org/10.1016/j.physrep.2004.08.031}{\emph{Phys. Rept.}
  {\bfseries 405} (2005) 279}
  [\href{https://arxiv.org/abs/hep-ph/0404175}{{\ttfamily hep-ph/0404175}}].

\bibitem{Feng:2010gw}
J.~L. Feng, \emph{{Dark Matter Candidates from Particle Physics and Methods of
  Detection}},
  \href{https://doi.org/10.1146/annurev-astro-082708-101659}{\emph{Ann. Rev.
  Astron. Astrophys.} {\bfseries 48} (2010) 495}
  [\href{https://arxiv.org/abs/1003.0904}{{\ttfamily 1003.0904}}].

\bibitem{Baer:2014eja}
H.~Baer, K.-Y. Choi, J.~E. Kim and L.~Roszkowski, \emph{{Dark matter production
  in the early Universe: beyond the thermal WIMP paradigm}},
  \href{https://doi.org/10.1016/j.physrep.2014.10.002}{\emph{Phys. Rept.}
  {\bfseries 555} (2015) 1} [\href{https://arxiv.org/abs/1407.0017}{{\ttfamily
  1407.0017}}].

\bibitem{Cirelli:2024ssz}
M.~Cirelli, A.~Strumia and J.~Zupan, \emph{{Dark Matter}},
  \href{https://arxiv.org/abs/2406.01705}{{\ttfamily 2406.01705}}.

\bibitem{Marsh:2015xka}
D.~J.~E. Marsh, \emph{{Axion Cosmology}},
  \href{https://doi.org/10.1016/j.physrep.2016.06.005}{\emph{Phys. Rept.}
  {\bfseries 643} (2016) 1} [\href{https://arxiv.org/abs/1510.07633}{{\ttfamily
  1510.07633}}].

\bibitem{DiLuzio:2020wdo}
L.~Di~Luzio, M.~Giannotti, E.~Nardi and L.~Visinelli, \emph{{The landscape of
  QCD axion models}},
  \href{https://doi.org/10.1016/j.physrep.2020.06.002}{\emph{Phys. Rept.}
  {\bfseries 870} (2020) 1} [\href{https://arxiv.org/abs/2003.01100}{{\ttfamily
  2003.01100}}].

\bibitem{Chadha-Day:2021szb}
F.~Chadha-Day, J.~Ellis and D.~J.~E. Marsh, \emph{{Axion dark matter: What is
  it and why now?}}, \href{https://doi.org/10.1126/sciadv.abj3618}{\emph{Sci.
  Adv.} {\bfseries 8} (2022) abj3618}
  [\href{https://arxiv.org/abs/2105.01406}{{\ttfamily 2105.01406}}].

\bibitem{Redondo:2008ec}
J.~Redondo and M.~Postma, \emph{{Massive hidden photons as lukewarm dark
  matter}}, \href{https://doi.org/10.1088/1475-7516/2009/02/005}{\emph{JCAP}
  {\bfseries 02} (2009) 005} [\href{https://arxiv.org/abs/0811.0326}{{\ttfamily
  0811.0326}}].

\bibitem{Jaeckel:2010ni}
J.~Jaeckel and A.~Ringwald, \emph{{The Low-Energy Frontier of Particle
  Physics}},
  \href{https://doi.org/10.1146/annurev.nucl.012809.104433}{\emph{Ann. Rev.
  Nucl. Part. Sci.} {\bfseries 60} (2010) 405}
  [\href{https://arxiv.org/abs/1002.0329}{{\ttfamily 1002.0329}}].

\bibitem{Oman:2015xda}
K.~A. Oman et~al., \emph{{The unexpected diversity of dwarf galaxy rotation
  curves}}, \href{https://doi.org/10.1093/mnras/stv1504}{\emph{Mon. Not. Roy.
  Astron. Soc.} {\bfseries 452} (2015) 3650}
  [\href{https://arxiv.org/abs/1504.01437}{{\ttfamily 1504.01437}}].

\bibitem{Luu:2024lfq}
H.~N. Luu et~al., \emph{{Diverse dark matter haloes in two-field fuzzy dark
  matter}}, \href{https://doi.org/10.1103/w9j1-k7b3}{\emph{Phys. Rev. D}
  {\bfseries 111} (2025) L121302}
  [\href{https://arxiv.org/abs/2408.00827}{{\ttfamily 2408.00827}}].

\bibitem{Amruth:2023xqj}
A.~Amruth et~al., \emph{{Einstein rings modulated by wavelike dark matter from
  anomalies in gravitationally lensed images}},
  \href{https://doi.org/10.1038/s41550-023-01943-9}{\emph{Nature Astron.}
  {\bfseries 7} (2023) 736} [\href{https://arxiv.org/abs/2304.09895}{{\ttfamily
  2304.09895}}].

\bibitem{2025arXiv250603132M}
J.~H. {Miller}, Jr and L.~L.~R. {Williams}, \emph{{A Fuzzy Situation Eased:
  Cold Dark Matter with Multipoles Can Explain The Double Radio Quad Lens HS
  0810+2554}}, \href{https://doi.org/10.48550/arXiv.2506.03132}{\emph{arXiv
  e-prints} (2025) arXiv:2506.03132}
  [\href{https://arxiv.org/abs/2506.03132}{{\ttfamily 2506.03132}}].

\bibitem{Bromley:2023yfi}
B.~C. Bromley, P.~Sandick and B.~Shams Es~Haghi, \emph{{Supermassive black hole
  binaries in ultralight dark matter}},
  \href{https://doi.org/10.1103/PhysRevD.110.023517}{\emph{Phys. Rev. D}
  {\bfseries 110} (2024) 023517}
  [\href{https://arxiv.org/abs/2311.18013}{{\ttfamily 2311.18013}}].

\bibitem{Koo:2023gfm}
H.~Koo, D.~Bak, I.~Park, S.~E. Hong and J.-W. Lee, \emph{{Final parsec problem
  of black hole mergers and ultralight dark matter}},
  \href{https://doi.org/10.1016/j.physletb.2024.138908}{\emph{Phys. Lett. B}
  {\bfseries 856} (2024) 138908}
  [\href{https://arxiv.org/abs/2311.03412}{{\ttfamily 2311.03412}}].

\bibitem{Koo:2025qac}
H.~Koo, \emph{{Head-on collisions of fuzzy/cold dark matter subhalos}},
  \href{https://doi.org/10.1007/s40042-025-01420-8}{\emph{J. Korean Phys. Soc.}
  {\bfseries 87} (2025) 430}
  [\href{https://arxiv.org/abs/2507.00607}{{\ttfamily 2507.00607}}].

\bibitem{Zhang:2021xxa}
H.-Y. Zhang, M.~Jain and M.~A. Amin, \emph{{Polarized vector oscillons}},
  \href{https://doi.org/10.1103/PhysRevD.105.096037}{\emph{Phys. Rev. D}
  {\bfseries 105} (2022) 096037}
  [\href{https://arxiv.org/abs/2111.08700}{{\ttfamily 2111.08700}}].

\bibitem{Jain:2021pnk}
M.~Jain and M.~A. Amin, \emph{{Polarized solitons in higher-spin wave dark
  matter}}, \href{https://doi.org/10.1103/PhysRevD.105.056019}{\emph{Phys. Rev.
  D} {\bfseries 105} (2022) 056019}
  [\href{https://arxiv.org/abs/2109.04892}{{\ttfamily 2109.04892}}].

\bibitem{Zhang:2023ktk}
H.-Y. Zhang, \emph{{Probing ultralight dark fields in cosmological and
  astrophysical systems}}, Ph.D. thesis, Rice U., 2023.
\newblock \href{https://arxiv.org/abs/2401.00043}{{\ttfamily 2401.00043}}.

\bibitem{Zhang:2024bjo}
H.-Y. Zhang, \emph{{Unified view of scalar and vector dark matter solitons}},
  \href{https://doi.org/10.1007/JHEP04(2025)174}{\emph{JHEP} {\bfseries 04}
  (2025) 174} [\href{https://arxiv.org/abs/2406.05031}{{\ttfamily
  2406.05031}}].

\bibitem{Amin:2022pzv}
M.~A. Amin, M.~Jain, R.~Karur and P.~Mocz, \emph{{Small-scale structure in
  vector dark matter}},
  \href{https://doi.org/10.1088/1475-7516/2022/08/014}{\emph{JCAP} {\bfseries
  08} (2022) 014} [\href{https://arxiv.org/abs/2203.11935}{{\ttfamily
  2203.11935}}].

\bibitem{Gorghetto:2022sue}
M.~Gorghetto, E.~Hardy, J.~March-Russell, N.~Song and S.~M. West, \emph{{Dark
  photon stars: formation and role as dark matter substructure}},
  \href{https://doi.org/10.1088/1475-7516/2022/08/018}{\emph{JCAP} {\bfseries
  08} (2022) 018} [\href{https://arxiv.org/abs/2203.10100}{{\ttfamily
  2203.10100}}].

\bibitem{Zhang:2023fhs}
H.-Y. Zhang and S.~Ling, \emph{{Phenomenology of wavelike vector dark matter
  nonminimally coupled to gravity}},
  \href{https://doi.org/10.1088/1475-7516/2023/07/055}{\emph{JCAP} {\bfseries
  07} (2023) 055} [\href{https://arxiv.org/abs/2305.03841}{{\ttfamily
  2305.03841}}].

\bibitem{Chen:2024vgh}
J.~Chen, L.~H. Nguyen and D.~J.~E. Marsh, \emph{{Vector dark matter halo: From
  polarization dynamics to direct detection}},
  \href{https://doi.org/10.1103/PhysRevD.111.043031}{\emph{Phys. Rev. D}
  {\bfseries 111} (2025) 043031}
  [\href{https://arxiv.org/abs/2407.17315}{{\ttfamily 2407.17315}}].

\bibitem{Amin:2023imi}
M.~A. Amin, A.~J. Long and E.~D. Schiappacasse, \emph{{Photons from dark photon
  solitons via parametric resonance}},
  \href{https://doi.org/10.1088/1475-7516/2023/05/015}{\emph{JCAP} {\bfseries
  05} (2023) 015} [\href{https://arxiv.org/abs/2301.11470}{{\ttfamily
  2301.11470}}].

\bibitem{Schiappacasse:2025mao}
E.~D. Schiappacasse, \emph{{Dark spin-2 field solitons as a source of
  electromagnetic radiation}},
  \href{https://doi.org/10.1088/1475-7516/2025/08/085}{\emph{JCAP} {\bfseries
  08} (2025) 085} [\href{https://arxiv.org/abs/2503.12569}{{\ttfamily
  2503.12569}}].

\bibitem{Caputo:2021eaa}
A.~Caputo, A.~J. Millar, C.~A.~J. O'Hare and E.~Vitagliano, \emph{{Dark photon
  limits: A handbook}},
  \href{https://doi.org/10.1103/PhysRevD.104.095029}{\emph{Phys. Rev. D}
  {\bfseries 104} (2021) 095029}
  [\href{https://arxiv.org/abs/2105.04565}{{\ttfamily 2105.04565}}].

\bibitem{Amaral:2024tjg}
D.~W.~P. Amaral, M.~Jain, M.~A. Amin and C.~Tunnell, \emph{{Vector wave dark
  matter and terrestrial quantum sensors}},
  \href{https://doi.org/10.1088/1475-7516/2024/06/050}{\emph{JCAP} {\bfseries
  06} (2024) 050} [\href{https://arxiv.org/abs/2403.02381}{{\ttfamily
  2403.02381}}].

\bibitem{Visinelli:2024wyw}
L.~Visinelli, T.~T. Yanagida and M.~Zantedeschi, \emph{{Do neutrinos bend?
  Consequences of an ultralight gauge field as dark matter}},
  \href{https://doi.org/10.1016/j.dark.2024.101659}{\emph{Phys. Dark Univ.}
  {\bfseries 46} (2024) 101659}
  [\href{https://arxiv.org/abs/2407.18300}{{\ttfamily 2407.18300}}].

\bibitem{Carenza:2025uwx}
P.~Carenza, T.~Ferreira and T.~T.~Q. Nguyen, \emph{{Out of the darkness:
  probing the inflationary era with dark photon dark matter}},
  \href{https://arxiv.org/abs/2507.08932}{{\ttfamily 2507.08932}}.

\bibitem{Amaral:2024rbj}
D.~W.~P. Amaral, D.~G. Uitenbroek, T.~H. Oosterkamp and C.~D. Tunnell,
  \emph{{First Search for Ultralight Dark Matter Using a Magnetically Levitated
  Particle}}, \href{https://doi.org/10.1103/PhysRevLett.134.251001}{\emph{Phys.
  Rev. Lett.} {\bfseries 134} (2025) 251001}
  [\href{https://arxiv.org/abs/2409.03814}{{\ttfamily 2409.03814}}].

\bibitem{Amaral:2025fcd}
D.~W.~P. Amaral, E.~D. Schiappacasse and H.-Y. Zhang, \emph{{Constraining Dark
  Photon Dark Matter with Radio Silence from Soliton Mergers around
  Supermassive Black Holes}},
  \href{https://arxiv.org/abs/2509.08932}{{\ttfamily 2509.08932}}.

\bibitem{Arias:2012az}
P.~Arias, D.~Cadamuro, M.~Goodsell, J.~Jaeckel, J.~Redondo and A.~Ringwald,
  \emph{{WISPy Cold Dark Matter}},
  \href{https://doi.org/10.1088/1475-7516/2012/06/013}{\emph{JCAP} {\bfseries
  06} (2012) 013} [\href{https://arxiv.org/abs/1201.5902}{{\ttfamily
  1201.5902}}].

\bibitem{Nakayama:2019rhg}
K.~Nakayama, \emph{{Vector Coherent Oscillation Dark Matter}},
  \href{https://doi.org/10.1088/1475-7516/2019/10/019}{\emph{JCAP} {\bfseries
  10} (2019) 019} [\href{https://arxiv.org/abs/1907.06243}{{\ttfamily
  1907.06243}}].

\bibitem{Mou:2022hqb}
Z.-G. Mou and H.-Y. Zhang, \emph{{Singularity Problem for Interacting Massive
  Vectors}}, \href{https://doi.org/10.1103/PhysRevLett.129.151101}{\emph{Phys.
  Rev. Lett.} {\bfseries 129} (2022) 151101}
  [\href{https://arxiv.org/abs/2204.11324}{{\ttfamily 2204.11324}}].

\bibitem{Nelson:2011sf}
A.~E. Nelson and J.~Scholtz, \emph{{Dark Light, Dark Matter and the
  Misalignment Mechanism}},
  \href{https://doi.org/10.1103/PhysRevD.84.103501}{\emph{Phys. Rev. D}
  {\bfseries 84} (2011) 103501}
  [\href{https://arxiv.org/abs/1105.2812}{{\ttfamily 1105.2812}}].

\bibitem{Graham:2015rva}
P.~W. Graham, J.~Mardon and S.~Rajendran, \emph{{Vector Dark Matter from
  Inflationary Fluctuations}},
  \href{https://doi.org/10.1103/PhysRevD.93.103520}{\emph{Phys. Rev. D}
  {\bfseries 93} (2016) 103520}
  [\href{https://arxiv.org/abs/1504.02102}{{\ttfamily 1504.02102}}].

\bibitem{Kolb:2020fwh}
E.~W. Kolb and A.~J. Long, \emph{{Completely dark photons from gravitational
  particle production during the inflationary era}},
  \href{https://doi.org/10.1007/JHEP03(2021)283}{\emph{JHEP} {\bfseries 03}
  (2021) 283} [\href{https://arxiv.org/abs/2009.03828}{{\ttfamily
  2009.03828}}].

\bibitem{Kolb:2023ydq}
E.~W. Kolb and A.~J. Long, \emph{{Cosmological gravitational particle
  production and its implications for cosmological relics}},
  \href{https://doi.org/10.1103/RevModPhys.96.045005}{\emph{Rev. Mod. Phys.}
  {\bfseries 96} (2024) 045005}
  [\href{https://arxiv.org/abs/2312.09042}{{\ttfamily 2312.09042}}].

\bibitem{Agrawal:2018vin}
P.~Agrawal, N.~Kitajima, M.~Reece, T.~Sekiguchi and F.~Takahashi, \emph{{Relic
  Abundance of Dark Photon Dark Matter}},
  \href{https://doi.org/10.1016/j.physletb.2019.135136}{\emph{Phys. Lett. B}
  {\bfseries 801} (2020) 135136}
  [\href{https://arxiv.org/abs/1810.07188}{{\ttfamily 1810.07188}}].

\bibitem{Co:2018lka}
R.~T. Co, A.~Pierce, Z.~Zhang and Y.~Zhao, \emph{{Dark Photon Dark Matter
  Produced by Axion Oscillations}},
  \href{https://doi.org/10.1103/PhysRevD.99.075002}{\emph{Phys. Rev. D}
  {\bfseries 99} (2019) 075002}
  [\href{https://arxiv.org/abs/1810.07196}{{\ttfamily 1810.07196}}].

\bibitem{Dror:2018pdh}
J.~A. Dror, K.~Harigaya and V.~Narayan, \emph{{Parametric Resonance Production
  of Ultralight Vector Dark Matter}},
  \href{https://doi.org/10.1103/PhysRevD.99.035036}{\emph{Phys. Rev. D}
  {\bfseries 99} (2019) 035036}
  [\href{https://arxiv.org/abs/1810.07195}{{\ttfamily 1810.07195}}].

\bibitem{Adshead:2023qiw}
P.~Adshead, K.~D. Lozanov and Z.~J. Weiner, \emph{{Dark photon dark matter from
  an oscillating dilaton}},
  \href{https://doi.org/10.1103/PhysRevD.107.083519}{\emph{Phys. Rev. D}
  {\bfseries 107} (2023) 083519}
  [\href{https://arxiv.org/abs/2301.07718}{{\ttfamily 2301.07718}}].

\bibitem{Kitajima:2023pby}
N.~Kitajima and F.~Takahashi, \emph{{Resonant production of dark photons from
  axions without a large coupling}},
  \href{https://doi.org/10.1103/PhysRevD.107.123518}{\emph{Phys. Rev. D}
  {\bfseries 107} (2023) 123518}
  [\href{https://arxiv.org/abs/2303.05492}{{\ttfamily 2303.05492}}].

\bibitem{Co:2021rhi}
R.~T. Co, K.~Harigaya and A.~Pierce, \emph{{Gravitational waves and dark photon
  dark matter from axion rotations}},
  \href{https://doi.org/10.1007/JHEP12(2021)099}{\emph{JHEP} {\bfseries 12}
  (2021) 099} [\href{https://arxiv.org/abs/2104.02077}{{\ttfamily
  2104.02077}}].

\bibitem{Long:2019lwl}
A.~J. Long and L.-T. Wang, \emph{{Dark Photon Dark Matter from a Network of
  Cosmic Strings}},
  \href{https://doi.org/10.1103/PhysRevD.99.063529}{\emph{Phys. Rev. D}
  {\bfseries 99} (2019) 063529}
  [\href{https://arxiv.org/abs/1901.03312}{{\ttfamily 1901.03312}}].

\bibitem{East:2022rsi}
W.~E. East and J.~Huang, \emph{{Dark photon vortex formation and dynamics}},
  \href{https://doi.org/10.1007/JHEP12(2022)089}{\emph{JHEP} {\bfseries 12}
  (2022) 089} [\href{https://arxiv.org/abs/2206.12432}{{\ttfamily
  2206.12432}}].

\bibitem{Cyncynates:2023zwj}
D.~Cyncynates and Z.~J. Weiner, \emph{{Detectable and Defect-Free Dark Photon
  Dark Matter}},
  \href{https://doi.org/10.1103/PhysRevLett.134.211002}{\emph{Phys. Rev. Lett.}
  {\bfseries 134} (2025) 211002}
  [\href{https://arxiv.org/abs/2310.18397}{{\ttfamily 2310.18397}}].

\bibitem{Cyncynates:2024yxm}
D.~Cyncynates and Z.~J. Weiner, \emph{{Experimental targets for dark photon
  dark matter}}, \href{https://doi.org/10.1103/PhysRevD.111.103535}{\emph{Phys.
  Rev. D} {\bfseries 111} (2025) 103535}
  [\href{https://arxiv.org/abs/2410.14774}{{\ttfamily 2410.14774}}].

\bibitem{Dong:2010in}
X.~Dong, B.~Horn, E.~Silverstein and A.~Westphal, \emph{{Simple exercises to
  flatten your potential}},
  \href{https://doi.org/10.1103/PhysRevD.84.026011}{\emph{Phys. Rev. D}
  {\bfseries 84} (2011) 026011}
  [\href{https://arxiv.org/abs/1011.4521}{{\ttfamily 1011.4521}}].

\bibitem{Arvanitaki:2019rax}
A.~Arvanitaki, S.~Dimopoulos, M.~Galanis, L.~Lehner, J.~O. Thompson and
  K.~Van~Tilburg, \emph{{Large-misalignment mechanism for the formation of
  compact axion structures: Signatures from the QCD axion to fuzzy dark
  matter}}, \href{https://doi.org/10.1103/PhysRevD.101.083014}{\emph{Phys. Rev.
  D} {\bfseries 101} (2020) 083014}
  [\href{https://arxiv.org/abs/1909.11665}{{\ttfamily 1909.11665}}].

\bibitem{Witten:1984dg}
E.~Witten, \emph{{Some Properties of O(32) Superstrings}},
  \href{https://doi.org/10.1016/0370-2693(84)90422-2}{\emph{Phys. Lett. B}
  {\bfseries 149} (1984) 351}.

\bibitem{Svrcek:2006yi}
P.~Svrcek and E.~Witten, \emph{{Axions In String Theory}},
  \href{https://doi.org/10.1088/1126-6708/2006/06/051}{\emph{JHEP} {\bfseries
  06} (2006) 051} [\href{https://arxiv.org/abs/hep-th/0605206}{{\ttfamily
  hep-th/0605206}}].

\bibitem{Arvanitaki:2009fg}
A.~Arvanitaki, S.~Dimopoulos, S.~Dubovsky, N.~Kaloper and J.~March-Russell,
  \emph{{String Axiverse}},
  \href{https://doi.org/10.1103/PhysRevD.81.123530}{\emph{Phys. Rev. D}
  {\bfseries 81} (2010) 123530}
  [\href{https://arxiv.org/abs/0905.4720}{{\ttfamily 0905.4720}}].

\bibitem{Arvanitaki:2010sy}
A.~Arvanitaki and S.~Dubovsky, \emph{{Exploring the String Axiverse with
  Precision Black Hole Physics}},
  \href{https://doi.org/10.1103/PhysRevD.83.044026}{\emph{Phys. Rev. D}
  {\bfseries 83} (2011) 044026}
  [\href{https://arxiv.org/abs/1004.3558}{{\ttfamily 1004.3558}}].

\bibitem{Dubovsky:2011tu}
S.~Dubovsky, A.~Lawrence and M.~M. Roberts, \emph{{Axion monodromy in a model
  of holographic gluodynamics}},
  \href{https://doi.org/10.1007/JHEP02(2012)053}{\emph{JHEP} {\bfseries 02}
  (2012) 053} [\href{https://arxiv.org/abs/1105.3740}{{\ttfamily 1105.3740}}].

\bibitem{McAllister:2014mpa}
L.~McAllister, E.~Silverstein, A.~Westphal and T.~Wrase, \emph{{The Powers of
  Monodromy}}, \href{https://doi.org/10.1007/JHEP09(2014)123}{\emph{JHEP}
  {\bfseries 09} (2014) 123} [\href{https://arxiv.org/abs/1405.3652}{{\ttfamily
  1405.3652}}].

\bibitem{Kaloper:2016fbr}
N.~Kaloper and A.~Lawrence, \emph{{London equation for monodromy inflation}},
  \href{https://doi.org/10.1103/PhysRevD.95.063526}{\emph{Phys. Rev. D}
  {\bfseries 95} (2017) 063526}
  [\href{https://arxiv.org/abs/1607.06105}{{\ttfamily 1607.06105}}].

\bibitem{Nomura:2017ehb}
Y.~Nomura, T.~Watari and M.~Yamazaki, \emph{{Pure Natural Inflation}},
  \href{https://doi.org/10.1016/j.physletb.2017.11.052}{\emph{Phys. Lett. B}
  {\bfseries 776} (2018) 227}
  [\href{https://arxiv.org/abs/1706.08522}{{\ttfamily 1706.08522}}].

\bibitem{Chatrchyan:2023cmz}
A.~Chatrchyan, C.~Er{\"o}ncel, M.~Koschnitzke and G.~Servant, \emph{{ALP dark
  matter with non-periodic potentials: parametric resonance, halo formation and
  gravitational signatures}},
  \href{https://doi.org/10.1088/1475-7516/2023/10/068}{\emph{JCAP} {\bfseries
  10} (2023) 068} [\href{https://arxiv.org/abs/2305.03756}{{\ttfamily
  2305.03756}}].

\bibitem{Peccei:1977hh}
R.~D. Peccei and H.~R. Quinn, \emph{{CP Conservation in the Presence of
  Instantons}}, \href{https://doi.org/10.1103/PhysRevLett.38.1440}{\emph{Phys.
  Rev. Lett.} {\bfseries 38} (1977) 1440}.

\bibitem{Weinberg:1977ma}
S.~Weinberg, \emph{{A New Light Boson?}},
  \href{https://doi.org/10.1103/PhysRevLett.40.223}{\emph{Phys. Rev. Lett.}
  {\bfseries 40} (1978) 223}.

\bibitem{Wilczek:1977pj}
F.~Wilczek, \emph{{Problem of Strong $P$ and $T$ Invariance in the Presence of
  Instantons}}, \href{https://doi.org/10.1103/PhysRevLett.40.279}{\emph{Phys.
  Rev. Lett.} {\bfseries 40} (1978) 279}.

\bibitem{Amin:2014eta}
M.~A. Amin, M.~P. Hertzberg, D.~I. Kaiser and J.~Karouby,
  \emph{{Nonperturbative Dynamics Of Reheating After Inflation: A Review}},
  \href{https://doi.org/10.1142/S0218271815300037}{\emph{Int. J. Mod. Phys. D}
  {\bfseries 24} (2014) 1530003}
  [\href{https://arxiv.org/abs/1410.3808}{{\ttfamily 1410.3808}}].

\bibitem{Amin:2011hj}
M.~A. Amin, R.~Easther, H.~Finkel, R.~Flauger and M.~P. Hertzberg,
  \emph{{Oscillons After Inflation}},
  \href{https://doi.org/10.1103/PhysRevLett.108.241302}{\emph{Phys. Rev. Lett.}
  {\bfseries 108} (2012) 241302}
  [\href{https://arxiv.org/abs/1106.3335}{{\ttfamily 1106.3335}}].

\bibitem{Lozanov:2016hid}
K.~D. Lozanov and M.~A. Amin, \emph{{Equation of State and Duration to
  Radiation Domination after Inflation}},
  \href{https://doi.org/10.1103/PhysRevLett.119.061301}{\emph{Phys. Rev. Lett.}
  {\bfseries 119} (2017) 061301}
  [\href{https://arxiv.org/abs/1608.01213}{{\ttfamily 1608.01213}}].

\bibitem{Lozanov:2014zfa}
K.~D. Lozanov and M.~A. Amin, \emph{{End of inflation, oscillons, and
  matter-antimatter asymmetry}},
  \href{https://doi.org/10.1103/PhysRevD.90.083528}{\emph{Phys. Rev. D}
  {\bfseries 90} (2014) 083528}
  [\href{https://arxiv.org/abs/1408.1811}{{\ttfamily 1408.1811}}].

\bibitem{Lozanov:2017hjm}
K.~D. Lozanov and M.~A. Amin, \emph{{Self-resonance after inflation: oscillons,
  transients and radiation domination}},
  \href{https://doi.org/10.1103/PhysRevD.97.023533}{\emph{Phys. Rev. D}
  {\bfseries 97} (2018) 023533}
  [\href{https://arxiv.org/abs/1710.06851}{{\ttfamily 1710.06851}}].

\bibitem{Copeland:1995fq}
E.~J. Copeland, M.~Gleiser and H.~R. Muller, \emph{{Oscillons: Resonant
  configurations during bubble collapse}},
  \href{https://doi.org/10.1103/PhysRevD.52.1920}{\emph{Phys. Rev. D}
  {\bfseries 52} (1995) 1920}
  [\href{https://arxiv.org/abs/hep-ph/9503217}{{\ttfamily hep-ph/9503217}}].

\bibitem{Salmi:2012ta}
P.~Salmi and M.~Hindmarsh, \emph{{Radiation and Relaxation of Oscillons}},
  \href{https://doi.org/10.1103/PhysRevD.85.085033}{\emph{Phys. Rev. D}
  {\bfseries 85} (2012) 085033}
  [\href{https://arxiv.org/abs/1201.1934}{{\ttfamily 1201.1934}}].

\bibitem{Zhang:2020bec}
H.-Y. Zhang, M.~A. Amin, E.~J. Copeland, P.~M. Saffin and K.~D. Lozanov,
  \emph{{Classical Decay Rates of Oscillons}},
  \href{https://doi.org/10.1088/1475-7516/2020/07/055}{\emph{JCAP} {\bfseries
  07} (2020) 055} [\href{https://arxiv.org/abs/2004.01202}{{\ttfamily
  2004.01202}}].

\bibitem{Zhang:2020ntm}
H.-Y. Zhang, \emph{{Gravitational effects on oscillon lifetimes}},
  \href{https://doi.org/10.1088/1475-7516/2021/03/102}{\emph{JCAP} {\bfseries
  03} (2021) 102} [\href{https://arxiv.org/abs/2011.11720}{{\ttfamily
  2011.11720}}].

\bibitem{Visinelli:2021uve}
L.~Visinelli, \emph{{Boson stars and oscillatons: A review}},
  \href{https://doi.org/10.1142/S0218271821300068}{\emph{Int. J. Mod. Phys. D}
  {\bfseries 30} (2021) 2130006}
  [\href{https://arxiv.org/abs/2109.05481}{{\ttfamily 2109.05481}}].

\bibitem{Visinelli:2017ooc}
L.~Visinelli, S.~Baum, J.~Redondo, K.~Freese and F.~Wilczek, \emph{{Dilute and
  dense axion stars}},
  \href{https://doi.org/10.1016/j.physletb.2017.12.010}{\emph{Phys. Lett. B}
  {\bfseries 777} (2018) 64}
  [\href{https://arxiv.org/abs/1710.08910}{{\ttfamily 1710.08910}}].

\bibitem{Hertzberg:2018zte}
M.~P. Hertzberg and E.~D. Schiappacasse, \emph{{Dark Matter Axion Clump
  Resonance of Photons}},
  \href{https://doi.org/10.1088/1475-7516/2018/11/004}{\emph{JCAP} {\bfseries
  11} (2018) 004} [\href{https://arxiv.org/abs/1805.00430}{{\ttfamily
  1805.00430}}].

\bibitem{Amin:2020vja}
M.~A. Amin and Z.-G. Mou, \emph{{Electromagnetic Bursts from Mergers of
  Oscillons in Axion-like Fields}},
  \href{https://doi.org/10.1088/1475-7516/2021/02/024}{\emph{JCAP} {\bfseries
  02} (2021) 024} [\href{https://arxiv.org/abs/2009.11337}{{\ttfamily
  2009.11337}}].

\bibitem{Planck:2018jri}
{\scshape Planck} collaboration, \emph{{Planck 2018 results. X. Constraints on
  inflation}}, \href{https://doi.org/10.1051/0004-6361/201833887}{\emph{Astron.
  Astrophys.} {\bfseries 641} (2020) A10}
  [\href{https://arxiv.org/abs/1807.06211}{{\ttfamily 1807.06211}}].

\bibitem{Kitajima:2021inh}
N.~Kitajima, K.~Kogai and Y.~Urakawa, \emph{{New scenario of QCD axion clump
  formation. Part I. Linear analysis}},
  \href{https://doi.org/10.1088/1475-7516/2022/03/039}{\emph{JCAP} {\bfseries
  03} (2022) 039} [\href{https://arxiv.org/abs/2111.05785}{{\ttfamily
  2111.05785}}].

\bibitem{Sikivie:2021trt}
P.~Sikivie and W.~Xue, \emph{{Resonant excitation of the axion field during the
  QCD phase transition}},
  \href{https://doi.org/10.1103/PhysRevD.105.043533}{\emph{Phys. Rev. D}
  {\bfseries 105} (2022) 043533}
  [\href{https://arxiv.org/abs/2110.13157}{{\ttfamily 2110.13157}}].

\bibitem{Kitajima:2018zco}
N.~Kitajima, J.~Soda and Y.~Urakawa, \emph{{Gravitational wave forest from
  string axiverse}},
  \href{https://doi.org/10.1088/1475-7516/2018/10/008}{\emph{JCAP} {\bfseries
  10} (2018) 008} [\href{https://arxiv.org/abs/1807.07037}{{\ttfamily
  1807.07037}}].

\bibitem{Garny:2018ali}
M.~Garny and J.~Heisig, \emph{{Interplay of super-WIMP and freeze-in production
  of dark matter}},
  \href{https://doi.org/10.1103/PhysRevD.98.095031}{\emph{Phys. Rev. D}
  {\bfseries 98} (2018) 095031}
  [\href{https://arxiv.org/abs/1809.10135}{{\ttfamily 1809.10135}}].

\bibitem{Baur:2017stq}
J.~Baur, N.~Palanque-Delabrouille, C.~Yeche, A.~Boyarsky, O.~Ruchayskiy,
  {\'E}.~Armengaud et~al., \emph{{Constraints from Ly-$\alpha$ forests on
  non-thermal dark matter including resonantly-produced sterile neutrinos}},
  \href{https://doi.org/10.1088/1475-7516/2017/12/013}{\emph{JCAP} {\bfseries
  12} (2017) 013} [\href{https://arxiv.org/abs/1706.03118}{{\ttfamily
  1706.03118}}].

\bibitem{Amin:2010jq}
M.~A. Amin and D.~Shirokoff, \emph{{Flat-top oscillons in an expanding
  universe}}, \href{https://doi.org/10.1103/PhysRevD.81.085045}{\emph{Phys.
  Rev. D} {\bfseries 81} (2010) 085045}
  [\href{https://arxiv.org/abs/1002.3380}{{\ttfamily 1002.3380}}].

\bibitem{Landau1976Mechanics}
L.~D. Landau and E.~M. Lifshitz, \emph{Mechanics, Third Edition: Volume 1
  (Course of Theoretical Physics)}. Butterworth-Heinemann, 3~ed., Jan., 1976.

\bibitem{Choi:2015fiu}
K.~Choi and S.~H. Im, \emph{{Realizing the relaxion from multiple axions and
  its UV completion with high scale supersymmetry}},
  \href{https://doi.org/10.1007/JHEP01(2016)149}{\emph{JHEP} {\bfseries 01}
  (2016) 149} [\href{https://arxiv.org/abs/1511.00132}{{\ttfamily
  1511.00132}}].

\bibitem{Higaki:2016yqk}
T.~Higaki, K.~S. Jeong, N.~Kitajima and F.~Takahashi, \emph{{Quality of the
  Peccei-Quinn symmetry in the Aligned QCD Axion and Cosmological
  Implications}}, \href{https://doi.org/10.1007/JHEP06(2016)150}{\emph{JHEP}
  {\bfseries 06} (2016) 150}
  [\href{https://arxiv.org/abs/1603.02090}{{\ttfamily 1603.02090}}].

\end{thebibliography}\endgroup
\end{document}